\documentclass[useAMS,usenatbib]{mn2e}

\usepackage[T1]{fontenc}
\usepackage{aecompl}
\usepackage{graphicx}
\usepackage{amssymb}
\usepackage{mathrsfs}
\usepackage[fleqn]{amsmath}
\usepackage{epstopdf,enumerate,url,color}

\usepackage{times}
\usepackage{flushend}

\title[Disc--planet eccentricities]{Growth of eccentric modes in disc--planet interactions}
\author[Teyssandier \& Ogilvie]{Jean Teyssandier\thanks{E-mail: jt591@cam.ac.uk} and Gordon I. Ogilvie\\ 
Department of Applied Mathematics and Theoretical Physics, University of Cambridge, Centre for Mathematical Sciences, Wilberforce Road,\\
Cambridge CB3 0WA, United Kingdom  
}

\newcommand{\pd}[2]{\frac{\partial #1}{\partial #2}}
\newcommand{\dd}[2]{\frac{\mathrm{d} #1}{\mathrm{d} #2}}
\newcommand{\id}{\mathop{}\!\mathrm{d}}
\newcommand{\im}{\mathrm{i}}
\newcommand{\me}{\mathrm{e}}
\newcommand{\cs}{c_\mathrm{s}^2}
\newcommand{\ab}{\alpha_\mathrm{b}}
\newcommand{\rin}{r_{\mathrm{in}}}
\newcommand{\rout}{r_{\mathrm{out}}}
\newcommand{\wout}{w_{\mathrm{out}}}
\newcommand{\rres}{r_{\mathrm{res}}}
\newcommand{\ee}{\left| E\right|^2}

\newcommand{\ff}{\mathcal{F}}
\newcommand{\gf}{\mathcal{G}}

\newcommand{\Mp}{M_{\rm p}}
\newcommand{\Md}{M_{\rm d}}
\newcommand{\qp}{q_{\rm p}}
\newcommand{\qd}{q_{\rm d}}
\newcommand{\Ep}{E_{\rm p}}
\newcommand{\ap}{a_{\rm p}}
\newcommand{\ep}{e_{\rm p}}
\newcommand{\Op}{\Omega_{\rm p}}

\newcommand{\ssc}{\Sigma_{\rm sc}}

\newcommand{\lc}[2]{b_{#1}^{(#2)}}

\begin{document}
\maketitle

\begin{abstract}
We formulate a set of linear equations that describe the behaviour of small eccentricities in a protoplanetary system consisting of a gaseous disc and a planet. Eccentricity propagates through the disc by means of pressure and self-gravity, and is exchanged with the planet via secular interactions. Excitation and damping of eccentricity can occur through Lindblad and corotation resonances, as well as viscosity. We compute normal modes of the coupled disc--planet system in the case of short-period giant planets orbiting inside an inner cavity, possibly carved by the stellar magnetosphere. Three-dimensional effects allow for a mode to be trapped in the inner parts of the disc. This mode can easily grow within the disc's lifetime. An eccentric mode dominated by the planet can also grow, although less rapidly. We compute the structure and growth rates of these modes and their dependence on the assumed properties of the disc.
\end{abstract}


\section{Introduction}
\label{sec:intro}

A significant fraction of extrasolar planets that have so far been discovered are known to have large orbital eccentricities. In particular, radial-velocity surveys have shown that giant planets far enough from their host stars not to suffer strong tidal circularization have eccentricities spanning the whole range from zero to near-unity \citep[see, e.g.,][]{marcy05,butler06}. \citet{kane12} confirmed that this feature was still present in the transit survey obtained with the \textit{Kepler} mission, and argued that the mean  eccentricity increases with planet size, although \cite{plavchan14} cautioned that uncertainties in the measurement of stellar radii in the \textit{Kepler} sample can affect this result. Theories of the formation and evolution of planetary systems have yet to explain this wide distribution of eccentricities.

It is likely that mutual gravitational interactions between planets in a gas-free environment can cause their eccentricities to grow to near-unity values. This can be the result of a violent scattering \citep{rf96,wm96,chatterjee08,jt08} or dynamical relaxation \citep{pt01,al03} of a large population of planets. Kozai-Lidov oscillations, caused by a stellar companion \citep{wm03,ft07,naoz12}, a planet \citep{naoz11}, or a disc \citep{ta10,ttp13}, can also pump up eccentricities. Finally, chaotic diffusion of angular momentum deficit can increase eccentricities and inclinations over very long timescales, a process known as secular chaos \citep{wl11}.  However, the initial conditions required to trigger these mechanisms often require the planets to have a certain amount of eccentricity to start with. It is unclear whether this amount is realistic or not, as it is currently difficult to predict planetary eccentricities at the beginning of the gas-free stage of dynamical evolution.

Another mechanism by which planets could gain eccentricities is their gravitational interaction with the protoplanetary disc in which they formed, in particular through mean-motion resonances between the planet and the disc. Since the resonant interactions involve an exchange of energy and angular momentum between the planet and disc, as well as a dissipative loss of energy from the system, they can affect the orbital eccentricities of both the planet and the disc.

It is now widely accepted that low-mass planets with small eccentricities that are unable to open a gap in a circular disc are expected to experience a relatively rapid eccentricity damping. \citet{ward88} showed that material orbiting at the same radius as the planet exerts a strong torque via coorbital resonances. This torque overcomes that of the non-coorbital resonances discussed by \citet{gt80} and results in eccentricity damping. This result holds as long as the surface density in the coorbital region is not strongly depleted. A refined analytical treatment of this problem was carried out by \citet{arty93}, and a similar result was obtained semi-analytically by \citet{tw04}. Numerical simulations of eccentric low-mass planets embedded in 2D and 3D discs \citep[see, e.g.,][]{cresswell07,bk10} have confirmed these findings. 

One the other hand, \citet{gt80} considered a satellite that is massive enough to open a gap in a circular disc (the same applies to gap-opening giant planets in protoplanetary discs), and showed that eccentric Lindblad resonances (ELRs) act to increase the eccentricity of the satellite, while the net effect of eccentric corotation resonances (ECRs) is to decrease the eccentricity. They concluded that, to lowest order in eccentricity, ECRs dominate by a small margin, and prevent eccentricity growth. An important breakthrough was achieved by \citet{gs03} and \citet{ol03} in their study of the saturation of the corotation torque \citep[see also][]{gt81}. \citet{gs03} noted that, unlike Lindblad resonances, corotation resonances readily undergo a nonlinear saturation. Thus there is a finite-amplitude instability which can occur if the eccentricity exceeds a small critical value. \citet{ol03} derived an expression for the level of saturation of the corotation torque in a three-dimensional gaseous disc. 

The above description makes the significant assumption that the disc remains circular. However, a general Keplerian disc involves nested elliptical orbits, and the secular interaction between the planet and the disc inevitably leads them to exchange eccentricity. Excitation of the eccentricity of a disc has been found to be important in various astrophysical contexts, such as in the rings of Uranus \citep{gt81}, or in superhump binary stars \citep{whitehurst88,lubow91}. A more general description is therefore desirable, in which both the planet and the disc are allowed to have eccentric orbits. Theories of eccentric gaseous discs have been discussed in the past by, e.g., \citet{kato83} and \citet{ogilvie01}. Moreover, \citet{ob14} recently developed a local model of eccentric discs, in which they explored non-linear vertical oscillations and deduced the linear global evolution of eccentricity, and then studied the hydrodynamical instability of such discs in \citet{bo14}.

Away from the coorbital region, the disc--planet interaction can be regarded as twofold. First, for the long-term orbit-averaged (secular) interaction, an analogy with planetary dynamics can be made. In celestial mechanics, the secular interaction between planets can be described by an eigenvalue problem, usually known as the Laplace--Lagrange theory \citep{md99}. In this theory, a reversible exchange of eccentricity occurs on long timescales (compared to orbital periods), and the total angular momentum deficit (AMD, a positive-definite measure of the eccentricity of the coupled system) of the system is conserved. (At this level of approximation, the inclinations undergo a similar, although decoupled, evolution.) A similar process can occur in disc--planet interactions, where the disc can be regarded as a continuum of rings of matter interacting with the planet, and with each other. As in the case of a planetary system, AMD is conserved in gravitational secular disc--planet interactions.
We remark here that for a planet on an initially eccentric orbit interacting with a circular disc, secular exchange of angular momentum would cause the eccentricity of the planet to decay at first, while that of the disc increases. 

The second part of the interaction occurs through mean-motion resonances, leading to an irreversible evolution in which the AMD can either increase or decrease. A theory of mean-motion resonances allowing for the eccentricities and inclinations of both the satellite and the disc to vary was formulated by \citet{ogilvie07}. Previously, a linear theory of disc--planet interactions was formulated by \citet{lo01} in the case of inclination, embodying secular and resonant contributions. Unlike previous work \citep[see, e.g.,][]{bgt84}, the disc was allowed to develop a warped shape, which is likely if the planet and the disc have comparable angular momenta. The linear theory allows for a set of normal modes which can grow or decay according to the balance between resonances (which mostly increase the AMD) and viscous dissipation of the warp (which decreases it). Unlike \citet{bgt84}, \citet{lo01} found that, after allowing the disc to become warped, inclination growth was suppressed by viscous damping for typical estimates of viscosity and disc parameters (although \citet{bgt84} were concerned with interaction between planetary rings and moons, while \citet{lo01} were studying protoplanetary discs). 

The results of \citet{lo01} regarding inclination highlight the fact that the secular evolution of the disc and that of the planet(s) are strongly coupled, and should be treated together rather than separately. The same statement can be made for eccentricities: in a system consisting of a disc and one or more planets, having comparable angular momenta, eccentricity is freely exchanged between the various components on secular timescales. To address the origin of eccentricity in protoplanetary systems, it is therefore more appropriate to ask under what conditions the total AMD of the coupled system, rather than the eccentricity of a planet, can grow.

Several numerical simulations of disc--planet interactions (or related systems) have investigated the possibility of eccentricity growth. \citet{arty91} found, using smoothed particle hydrodynamics (SPH), that the eccentricity of a binary star can increase through its interaction with a circumbinary disc. In this case the gap is so wide that only the 1:3 ELR is present (and no ECR), so that the damping argument of \citet{gt80} is not applicable. The mechanism is analogous to that operating in superhump binary stars, where the eccentricity of the circumstellar disc is excited and only the 3:1 resonance is operative. Such excitation was readily observed in grid-based simulations by \citet{kley08}, with rapid growth rates. In the case of disc--planet interactions, \citet{papaloizou01} showed that massive planets (e.g., those of 20 Jupiter masses) could experience eccentricity growth, with the exterior disc also becoming noticeably eccentric. \citet{kd06} found that, even if the planet is held on a fixed circular orbit, the disc can become eccentric for planets exceeding about 3 Jupiter masses (for typical disc parameters). Without fixing the planet's orbit, \citet{dangelo06} found growing eccentricities for planetary masses as small as 1 Jupiter mass. This extension of the work by \citet{arty91} and \citet{papaloizou01} was made possible through the use of high-resolution two-dimensional numerical simulations over several thousand planetary orbits. In contrast with circumbinary discs, they found that the 1:3 ELR was highly ineffective at pumping the eccentricity, and that the 2:4 and 3:5 ELRs were the most efficient ones, with additional contributions arising from resonances located within the gap. \citet{cresswell07} studied the evolution of eccentricity and inclination of non-gap-opening planets, and confirmed that their eccentricities (and inclinations) would be damped by the disc, in good agreement with the semi-analytical work by \citet{tw04}. This result was later confirmed by \citet{bk10} for various planet masses, up to 1 Jupiter mass, in three-dimensional isothermal and fully radiative discs. Perhaps most relevant to the present work is the study by \citet{rice08} of a hot Jupiter orbiting inside the magnetospheric cavity of the star. The authors found eccentricity growth for massive planets, which would eventually lead to the destruction of the planet, and possibly explain the paucity of very massive planets on short-period orbits. We will discuss these claims in the present paper. More recently, \citet{dunhill13} have claimed, using SPH simulations, that only very massive planets (e.g., those of $20$ Jupiter masses) embedded in discs with high surface densities could experience eccentricity growth, although their simulations were limited to a few hundred orbital periods. One the other hand, \citet{dc15} recently argued that disc--planet interactions can indeed increase the eccentricity of a Jupiter-mass planet embedded in an isothermal disc. Such growth is possible when the planet opens a deep gap, and only if its initial eccentricity exceeds a threshold value, similarly to what was predicted analytically by \citet{gs03}. Finally, \citet{tsang14} argued in favour of eccentricity growth for gap-opening planets in non-barotropic discs. They showed that when stellar illumination heats the gap to a certain threshold, the corotation torque can be reduced to a point where it is not able to damp eccentricity.

In this paper we present a linear theory of eccentricity evolution in disc--planet systems. Our theory is analogous to the one established for inclination by \citet{lo01}. It is also related to the work by \citet{papaloizou02} but differs by including a new treatment of mean-motion resonances between planets and discs \citep{ogilvie07} and a three-dimensional description of the dynamics of eccentric discs \citep{ogilvie01,ogilvie08}. We aim to compare our results with previous analytical studies \citep[e.g.,][]{gs03} as well as numerical simulations such as that of \citet{rice08}. The equations describing the behaviour of eccentricity are presented in Section \ref{sec:evo_eqs}. The associated conservation laws and equations for the angular momentum deficit of the system are obtained in Section \ref{sec:amdint}, together with integral relations for the precession rates and the growth rates of normal modes. An analogy with the Schr\"odinger equation and an application to a numerical solution for a non-self-gravitating disc are presented in Section \ref{sec:fidu}, while a more complete study that includes additional physics is carried out in Section \ref{sec:fiduall}. In Section \ref{sec:param} we conduct a numerical exploration of the influence of various physical parameters. In Section \ref{sec:discu} we discuss our results, and we conclude in Section \ref{sec:conclu}.

\section{Evolutionary equations for an eccentric disc--planet system}
\label{sec:evo_eqs}

In this section we present a set of linear equations that describe the evolution of small eccentricities in a disc--planet system. To a first approximation, fluid elements in an eccentric disc follow elliptical Keplerian orbits. The eccentricity $e$ and argument of pericenter $\varpi$ can vary smoothly with the orbital semi-major axis $a$. For small gradients in $e$ and $\varpi$, the orbits are nested without intersections \citep{ogilvie01}. 

Throughout this paper the results are presented in terms of the complex eccentricity $E=e\,\me^{\im \varpi}$. Since $|E|=e$ and $\arg{(E)}=\varpi$, this variable conveniently describes both the shape and the orientation of the elliptical orbits. With similar notations, the planet has a complex eccentricity $\Ep =\ep\,\me^{\im \varpi_{\rm p}} $. We also denote by $\ap$, $\Op$ and $\Mp$ the semi-major axis, orbital frequency and mass of the planet, with $\qp=\Mp/M_*$ the planet-to-star mass ratio. In addition we denote by $\Md$ the total mass of the disc and $\qd=\Md/M_*$.

\subsection{Pressure}
\label{sec:pres}

\citet{go06} derived a linear partial differential equation governing the propagation of a small complex eccentricity $E(r,t)$ in a two-dimensional, non-self-gravitating, adiabatic disc. It has the form of a dispersive wave equation related to the Schr\"odinger equation. In Appendix \ref{app:lineareq} we derive a similar equation for a three-dimensional disc in which the isothermal sound speed $c_{\rm s}$ is a specified function of radius, as is commonly assumed in numerical simulations of disc--planet systems. A similar method can be used to derive an equation for a  3D adiabatic disc \citep[see also][]{ob14}. The different models take the following forms:
\begin{itemize}
\item 2D adiabatic model:
\begin{equation}
\label{eq:2dadia}
\Sigma r^2 \Omega \pd{E}{t} = \frac{\im}{r}\pd{}{r}\left( \frac{1}{2}\gamma Pr^3 \pd{E}{r} \right) + \frac{\im r}{2}\dd{P}{r}E.
\end{equation}
\item 3D adiabatic model:
\begin{align}
\label{eq:3dadia}
\Sigma r^2 \Omega \pd{E}{t} &= \frac{\im}{r}\pd{}{r}\left[ \frac{1}{2}\left(2-\frac{1}{\gamma}\right)Pr^3 \pd{E}{r} \right]\nonumber\\
& + \frac{\im}{2}\left(4-\frac{3}{\gamma}\right)r\dd{P}{r}E + \frac{3\im}{2}\left(1+\frac{1}{\gamma}\right)PE.
\end{align}
\item 2D locally isothermal model:
\begin{align}
\label{eq:2diso}
\Sigma r^2 \Omega \pd{E}{t} &= \frac{\im}{r}\pd{}{r}\left( \frac{1}{2}\Sigma \cs r^3 \pd{E}{r} \right) + \frac{\im r}{2}\dd{}{r}\left(\Sigma\cs \right)E\nonumber\\
& - \frac{\im}{2r}\pd{}{r}\left(\Sigma \dd{\cs}{r} r^3 E\right).
\end{align}
\item 3D locally isothermal model:
\begin{align}
\label{eq:3diso}
\Sigma r^2 \Omega \pd{E}{t} &= \frac{\im}{r}\pd{}{r}\left( \frac{1}{2}\Sigma \cs r^3 \pd{E}{r} \right) + \frac{\im r}{2}\dd{}{r}\left(\Sigma\cs \right)E\nonumber\\
& - \frac{\im}{2r}\pd{}{r}\left(\Sigma \dd{\cs}{r} r^3 E\right) + \frac{3\im}{2r}\Sigma\dd{}{r}\left(\cs r^2\right)E.
\end{align}
\end{itemize}
Here $\Omega = (GM_*/r^3)^{1/2}$ is the Keplerian angular velocity around a star of mass $M_*$, although our equations do take into account the small departure from Keplerian rotation arising from the radial pressure gradient. In addition, $\Sigma$ is the surface density, $P$ is the vertically integrated pressure, and $\gamma$ is the adiabatic index. (In the 2D model, $\Sigma$ and $P$ are simply the density and pressure that appear in the two-dimensional fluid-dynamical equations.) Note that $P=\Sigma\cs$ in a locally isothermal disc. There is an equivalence between the adiabatic and isothermal models when setting $\gamma=1$ and assuming that $c_{\rm s}$ is independent of $r$. 

Differences between the 2D and 3D models arise from the periodic variation of the vertical gravitational force exerted on a fluid element as it goes along its elliptical orbit. The vertical variation of the radial gravitational force also makes a difference. These effects account for the last term in each of the 3D equations, as well as for the different coefficients in the adiabatic case. In Eq. (\ref{eq:3diso}) the last term derives from 3D effects, while the preceding `non-adiabatic' term results from the variation of the isothermal sound speed with radius. The very significant role of the 3D term for eccentric discs around Be stars has already been demonstrated by \citet{ogilvie08}.

\subsection{Self-gravity and secular interactions with a companion}

The self-gravity of the disc can be in treated in a secular way as a continuum version of the classical Laplace--Lagrange theory \citep[see, e.g.,][]{md99}. In Appendix \ref{app:pertpot}, we show that self-gravity gives rise to the following contribution:
\begin{align}
\label{eq:esg}
\lefteqn{\Sigma r^2 \Omega \left( \pd{E}{t} \right)_{\rm dd}= \int \im G\Sigma(r)\Sigma(r')}&\nonumber\\
&&\times\left[K_1(r,r')E(r,t) - K_2(r,r')E(r',t)\right]2\pi r' \,\id r',
\end{align}
where
\begin{equation}
K_m(r,r') = \frac{rr'}{4\pi}\int_{0}^{2\pi}\frac{\cos m\theta \,\id \theta}{\left(r^2+r'^2-2rr'\cos\theta\right)^{3/2}}
\end{equation}
is a symmetric kernel representing the strength of the gravitational interaction between slightly elliptical rings at radii $r$ and $r'$. In equation (\ref{eq:esg}) the term $K_1(r,r')$ represents the contribution to the precession of the annulus at $r$ arising from the axisymmetric gravitational potential generated by the annulus at $r'$, while $K_2(r,r')$ represents the forcing of the eccentricity of the first annulus by that of the second.  We resolve the singularity at $r=r'$ by replacing $K_m$ with a softened symmetric kernel:
\begin{align}
K_m(r,r') &= \frac{rr'}{4\pi}\int_{0}^{2\pi}\frac{\cos m\theta \,\id \theta}{\left( r^2+r'^2-2rr'\cos\theta + s^2rr' \right)^{3/2}}  \nonumber\\
&= \frac{\beta^{3/2}}{4(rr')^{1/2}}b_{3/2}^{(m)}(\beta),
\end{align}
where $s$ is a dimensionless softening parameter, $b$ is a Laplace coefficient,\begin{equation}
\label{eq:lc}
b_{k}^{(m)}(\beta) = \frac{1}{\pi}\int_0^{2\pi} \frac{\cos m \theta \,\id \theta}{(1-2\beta \cos \theta 	+ \beta^2)^k},
\end{equation}
and $\beta$ is the smaller solution ($\beta<1$) of the quadratic equation
\begin{equation}
\frac{1+\beta^2}{\beta} = \frac{r^2+r'^2}{rr'} + s^2.
\end{equation}
The softening length is then $s(rr')^{1/2}$, and, in the limit $s\to 0$, $\beta \to \min(r,r')/\max(r,r')$. We set $s$ to be the scaling factor $h_0$ which appears in the expression for the disc aspect ratio (see the description of the disc model in Section \ref{sec:discmodel}). This gives an approximation of the averaging of the gravitational potential of the disc over its vertical extent. 

Self-gravity is therefore an additional means of propagation of eccentricity, as it gives rise to a global coupling of eccentricity between different parts of the disc. The role of self-gravity in eccentric discs was already discussed by \citet{tremaine01}.

The secular potential of a planet can be regarded as that of an elliptical ring whose mass is that of the planet. Hence its contribution to the eccentricity evolution is not fundamentally different from that of self-gravity:
\begin{align}
\label{eq:epd1}
\Sigma r^2 \Omega & \left( \pd{E}{t}\right)_{\rm pd} = \im G\Mp\Sigma(r)\nonumber\\
&\times\left[K_1(r,\ap)E(r,t) - K_2(r,\ap)\Ep\right].
\end{align}
Likewise, the contribution to the eccentricity evolution of the planet is
\begin{align}
\label{eq:epd2}
\Mp \ap^2 \Op & \left(\dd{\Ep}{t}\right)_{\rm pd} = \int\im G\Mp\Sigma(r)\nonumber\\
&\times\left[ K_1(\ap,r)\Ep - K_2(\ap,r) E(r,t) \right]2\pi r \,\id r.
\end{align}
We note that in the linear theory of \citet{lo01}, the gravitational couplings in disc--planet systems lead to identical equations for the complex inclination, except for a change of sign and the replacement of $K_2$ with $K_1$. It is straightforward to generalize the equations to the case of a multiple-planet system.

\subsection{Viscosity}
 The mechanism by which angular momentum is transported in protoplanetary discs remains poorly understood. Even if this is modelled as an effective shear viscosity, its influence on the eccentricity is complicated and can even lead to growth rather than decay in some circumstances \citep{lo06}.
In the present we adopt a simplified model of eccentricity damping. Following \citet{go06}, we employ an effective bulk viscosity described by a Shakura-Sunyaev $\alpha$-parametrization, which leads to
\begin{equation}
\Sigma r^2 \Omega \left( \pd{E}{t} \right)_{\rm visc} = \frac{1}{2r}\pd{}{r}\left(\ab Pr^3 \pd{E}{r} \right),
\end{equation}
where $\ab$ is a dimensionless parameter such that the effective bulk viscosity is given in terms of the pressure $p$ by $\ab p/\Omega$. This diffusive term reduces the AMD of the system, and is therefore a convenient representation of eccentricity damping. This effective bulk viscosity represents whichever process (thermal or mechanical), apart from resonances, acts to damp the eccentricity, and should not necessarily be regarded as a hydrodynamic viscosity.

\subsection{Resonances}
\label{sec:res}

When the planet is on an eccentric orbit, its gravitational potential can be expanded in a Fourier series of the form
\begin{equation}
\Psi(r,\phi,t) = \sum_{l,m}\psi_{l,m}(r)\exp{[\im(m\phi-l\Op t)]}
\end{equation}
where the $\psi_{l,m}$ coefficients can be found, e.g., in \citet{gt78}. Here we simply state that they are proportional to $e_\mathrm{p}^{|l-m|}$ where $e_\mathrm{p}\ll1$ is the planet's eccentricity. For a planet on a circular orbit, only terms with $l=m$ appear. At the first order in eccentricity, we keep terms with $l=m\pm 1$. Here and in the remaining of Section \ref{sec:res} (and also in Section \ref{sec:amdint}), we adopt the convention that the upper sign applies to inner resonances (i.e., a disc interior to the planet) and the lower sign applies to outer resonances (i.e., a disc exterior to the planet).

In a linear theory in which we work to first order in $e$, two types of resonances need to be considered. Eccentric Lindblad resonances (ELRs) correspond to locations in the disc where the perturbing frequency in the rotating frame $l\Omega_p - m\Omega$ matches $\mp\kappa$, the epicyclic frequency . We further assume a Keplerian disc, for which $\kappa=\Omega$. Hence, inner (resp. outer) ELRs are of the form $\Omega/\Op=j/(j-2)$ (resp. $\Omega/\Op=(j-2)/j$) for $j\geq3$, where $j=m+1$. 
Eccentric corotation resonances (ECRs) correspond to $l\Op - m\Omega = 0$. Hence they occur at commensurabilities of the form $\Omega/\Op=j/(j-1)$ and $\Omega/\Op=(j-1)/j$ for $j\geq2$, with $j=m+1$ for an inner ECR and $j=m$ for an outer ECR, respectively. Irreversible growth and decay of eccentricity may occur because of resonant interactions between the planet and the disc. 

\subsubsection{Eccentric Lindblad resonances}

As we show in Appendix \ref{app:res}, based on \citet{ogilvie07}, a single ELR will contribute to the equation for the complex eccentricities of the disc and planet in the following way: 
\begin{align}
\label{eq:elrd}
\Sigma r^2 \Omega & \left(\pd{E}{t}\right)_{\rm ELR}  = \frac{G\Mp^2}{M_*}\Sigma\mathscr{A} \left(\mathscr{A}E-\mathscr{B}\Ep \right)\nonumber\\
&\times w_{\rm L}^{-1}\Delta\left( \frac{r-\rres}{w_{\rm L}} \pm 1 \right),
\end{align}
\begin{align}
\label{eq:elrp}
\Mp \ap^2 \Op & \left( \dd{\Ep}{t} \right)_{\rm ELR} = \frac{G\Mp^2}{M_*}\int\Sigma\mathscr{B} \left(\mathscr{B}\Ep-\mathscr{A}E \right)\nonumber\\
&\times w_{\rm L}^{-1}\Delta\left( \frac{r-\rres}{w_{\rm L}} \pm 1 \right) 2\pi r \,\id r.
\end{align}
\begin{table}
  \begin{tabular}{cccc}
    $j$ &  $\rres/\ap$ & $\mathscr{A}$ & $\mathscr{B}$ \\ \hline
3 &     0.4807 &     0.8332 &     1.5397 \\
4 &     0.6300 &     2.1863 &     3.2019 \\
5 &     0.7114 &     3.8918 &     5.1463 \\
6 &     0.7631 &     5.8890 &     7.3456 \\
7 &     0.7991 &     8.1406 &     9.7756 \\
8 &     0.8255 &    10.6210 &    12.4173 \\
9 &     0.8457 &    13.3108 &    15.2555 \\
10 &     0.8618 &    16.1950 &    18.2777 \\
$\cdots$ & $\cdots$ & $\cdots$ & $\cdots$ \\
10 &     1.1604 &    21.3831 &    18.9703 \\
9 &     1.1824 &    18.2033 &    15.9081 \\
8 &     1.2114 &    15.1989 &    13.0272 \\
7 &     1.2515 &    12.3809 &    10.3395 \\
6 &     1.3104 &     9.7628 &     7.8592 \\
5 &     1.4057 &     7.3617 &     5.6037 \\
4 &     1.5874 &     5.2009 &     3.5940 \\
3 &     2.0801 &     0.6072 &     1.8490 \\
\hline
  \end{tabular}
  \caption{Resonant radii and coefficients for eccentric Lindblad resonances}
  \label{tab:rab}
\end{table}
The resonant radii $\rres$ and the dimensionless coefficients $\mathscr{A}$ and $\mathscr{B}$ are displayed in Table \ref{tab:rab}. These equations show that a single ELR tends to cause an exponential growth of either the eccentricity of the disc or that of the planet, if the other is circular. In general it is a `relative' eccentricity, as measured by the linear combination of $\mathscr{A}E-\mathscr{B}\Ep$, that tends to grow. The radial profile of the resonance is described through the dimensionless function $\Delta$ and a resonant width $w_{\rm L}$, which represents the broadening of Lindblad resonances by collective effects such as pressure and self-gravity. \citet{mvs87} showed that the radial distribution of the torque exerted at a Lindblad resonance in a non-self-gravitating disc is given by an Airy function of a scaled radial coordinate. The peak of the Airy function $\mathrm{Ai}(x)$ is very close to $x=-1$, which is on the side of the resonance on which a density wave is emitted. We model the torque distribution using a Gaussian function $\Delta(x)=(2\pi)^{-1/2}\exp(-x^2/2)$ that is off-centred by one resonant width.  We find that a Gaussian evaluated at $((r-\rres)/w_{\rm L}) \pm 1$ gives a reasonable representation of the Airy function. We recall that, following our convention, the $+$ (resp. $-$) sign is for an inner (resp. outer) Lindblad resonance. 
As we will discuss further, the off-centred torque distribution has the important effect of pushing more resonances inside the disc. 

A simple estimate of the resonant width $w_{\rm L}$ can found from the dispersion relation for density waves in a Keplerian disc ($\kappa=\Omega$), where self-gravity is neglected:
\begin{equation}
\label{eq:disprel}
(l\Op - m\Omega)^2 = \Omega^2 + \cs k_r^2.
\end{equation}
Here $k_r$ is the radial wavenumber, which vanishes at the resonance. Lindblad resonances launch a wave in the disc, which propagates away from the planet. Differentiating equation (\ref{eq:disprel}) gives
\begin{equation}
-2m(l\Op-m\Omega)\,\id\Omega = 2\Omega\,\id\Omega + \cs\,\id k_r^2.
\end{equation}
Substituting $l\Op-m\Omega$ by $\mp\Omega$ and assuming that $\id k_r^2/\id r \approx \mp w_{\rm L}^{-3}$, which means that the wavenumber has increased to $w_{\rm L}^{-1}$ at a distance $w_{\rm L}$ from the resonance, we find
\begin{equation}
\label{eq:rwelr}
\frac{w_{\rm L}}{r}\bigg|_{\rm ELR} \approx \left( \frac{h_0^2}{3(m\mp 1)} \right)^{1/3},
\end{equation}
where $h_0$ is a typical value of the aspect ratio of the disc (see Section \ref{sec:discmodel}). Indeed, \citet{mvs87} already showed that the width of Lindblad resonances in a non-self-gravitating disc was proportional to $(H/r)^{2/3}$. \citet{lubow10} found that the non-zero width of the 3:1 ELR was important for the eccentric mode in superhump binaries. Table \ref{tab:rab} shows that resonances with large values of $j$ are intrinsically stronger than those with low values of $j$, but accumulate close to the planet, where the surface density is likely to be zero. The dependence of equation (\ref{eq:rwelr}) on $m=j-1$ means that resonances with large values of $j$ have a decreasing width. In the remainder of this paper, we apply a torque cutoff at $j\gtrsim r/H$ which ensures that resonances with higher $j$ are ineffective \citep{gt80}. Finally, we note that the coefficient $\mathscr{A}$ is greatly reduced for the 1:3 resonance, because of the contribution of the indirect term in equation (\ref{eq:appab}). Hence we expect this resonance to give only a small contribution to the eccentricity growth in the disc, especially if the planet is fixed on a circular orbit.

\subsubsection{Eccentric corotation resonances}

Similarly to what we derived for ELRs, a single ECR gives the following contribution:
\begin{align}
\label{eq:ecrd}
\Sigma r^2 \Omega & \left( \pd{E}{t} \right)_{\rm ECR} = \pm\dd{\ln (\Sigma/\Omega)}{\ln r}\frac{G\Mp^2}{M_*}\Sigma\mathscr{C} \left(\mathscr{C}E-\mathscr{D}\Ep \right)\nonumber\\
&\times w_{\rm C}^{-1}\Delta\left( \frac{r-\rres}{w_{\rm C}} \right),
\end{align}
\begin{align}
\label{eq:ecrp}
\Mp \ap^2 \Op & \left( \dd{\Ep}{t} \right)_{\rm ECR} = \pm\dd{\ln (\Sigma/\Omega)}{\ln r} \frac{G\Mp^2}{M_*}\int\Sigma\mathscr{D} \left(\mathscr{D}\Ep-\mathscr{C}E \right)\nonumber\\
&\times w_{\rm C}^{-1}\Delta\left( \frac{r-\rres}{w_{\rm C}} \right) 2\pi r \,\id r.
\end{align}
where again, the $+$ (resp. $-$) sign is for an inner (resp. outer) corotation resonance.
\begin{table}
  \begin{tabular}{cccc}
    $j$ &  $\rres/\ap$ & $\mathscr{C}$ & $\mathscr{D}$ \\ \hline
2 &     0.6300 &     1.0853 &     0.3906 \\
3 &     0.7631 &     2.2367 &     2.7434 \\
4 &     0.8255 &     3.3933 &     3.9223 \\
5 &     0.8618 &     4.5517 &     5.0930 \\
6 &     0.8855 &     5.7109 &     6.2601 \\
7 &     0.9023 &     6.8705 &     7.4252 \\
8 &     0.9148 &     8.0304 &     8.5890 \\
9 &     0.9245 &     9.1905 &     9.7522 \\
$\cdots$ & $\cdots$ & $\cdots$ & $\cdots$ \\
9 &     1.0817 &     9.9412 &    10.5488 \\
8 &     1.0931 &     8.7780 &     9.3887 \\
7 &     1.1082 &     7.6141 &     8.2288 \\
6 &     1.1292 &     6.4489 &     7.0692 \\
5 &     1.1604 &     5.2817 &     5.9099 \\
4 &     1.2114 &     4.1107 &     4.7515 \\
3 &     1.3104 &     2.9309 &     3.5949 \\
2 &     1.5874 &     1.7229 &     0.6200 \\
\hline
  \end{tabular}
  \caption{Resonant radii and coefficients for eccentric corotation resonances}
  \label{tab:rcd}
\end{table}
Fluid elements in the corotation region are subject to libration around the nominal resonant radius. In the absence of dissipative effects such as viscosity, the corotation region is a closed system with a limited budget of angular momentum to transfer to the planet. Hence, the corotation torque will eventually saturate in a non-dissipative disc. When viscosity is included, the torque will also eventually vanish if the viscous diffusion timescale across the libration region is larger than the libration timescale. Thus, there exists a critical viscosity below which eccentricity growth is possible. In this picture, the resonant width is defined by the size of the libration zone. An estimate of this width was obtained by \citet{gt81} using a ballistic treatment.
 
Following \citet{mo04}, we adopt the following form:
\begin{equation}
\label{eq:rwecr}
\frac{w_{\rm C}}{r}\bigg|_{\rm ECR} \approx  4.1\left( C_m^{\pm}me\qp \right )^{1/2},
\end{equation}
where $C_m^{\pm}$ are coefficients which can be found in Tables 1 and 2 of \cite{ol03}, and which we set to unity for simplicity, as most of the uncertainty in our knowledge of the width lies in our choice for $e$ (where $e$ could be the eccentricity of the planet, that of the disc, or a linear combination of the two). The formula of \citet{mo04} gives the full width of the resonance, and thus should be divided by a factor of $\sim 5$ to give the correct Gaussian width. In practice, this means that the width of ECRs can be significantly smaller than the width of ELRs. 

\subsection{Short-range forces}

\subsubsection{General relativity correction}

Relativistic corrections at the first order in post-Newtonian expansion cause an additional, prograde, contribution to the apsidal motion of the planet and the disc. Its effect has long been incorporated and appreciated in the framework of secular planetary dynamics \citep[see, e.g.,][]{al06}. To leading order in eccentricity, it reads
\begin{equation}
\label{eq:grd}
\Sigma r^2 \Omega \left( \pd{E}{t} \right)_{\rm 1PN} = 3\im\frac{G^2M_*^2}{c^2r^2} \Sigma(r)E(r,t)
\end{equation}
for the eccentricity in the disc, and
\begin{equation}
\label{eq:grp}
\Mp \ap^2 \Op \left( \dd{\Ep}{t} \right)_{\rm 1PN} = 3\im\frac{G^2M_*^2}{c^2\ap^2} \Mp \Ep
\end{equation}
for the eccentricity of the planet (here $c$ is the speed of light in vacuum).

\subsubsection{Stellar oblateness}

The stellar rotation causes the star to become oblate, generating a force on the orbit that causes it to progress in a prograde manner. This can be easily included in the framework of secular theory \citep[see, e.g.,][]{md99} as
\begin{equation}
\label{eq:j2d}
\Sigma r^2 \Omega \left( \pd{E}{t} \right)_{\rm J_2} = \frac{3\im}{2}\frac{GM_*}{r} \Sigma(r) J_2\left(\frac{R_*}{r}\right)^2E(r,t)
\end{equation}
and
\begin{equation}
\label{eq:j2p}
\Mp \ap^2 \Op \left( \dd{\Ep}{t} \right)_{\rm J_2} = \frac{3\im}{2}\frac{GM_*}{\ap} \Mp J_2\left(\frac{R_*}{\ap}\right)^2 \Ep,
\end{equation}
where $R_*$ is the stellar radius. In addition $J_2$ represents the leading order correction from a spherically symmetric gravitational potential. Values of $J_2$ are poorly known and depend on the stellar spin frequency. Values for the Sun are typically low (about $2\times 10^{-6}$), but could be higher for fast-rotating T Tauri-type stars. 

Although we give the expression for the stellar oblateness, we have not included it in any of the calculations in the remaining of the paper. Here we merely state that for rapidly rotating stars, its influence on the precession of a hot Jupiter could be as important as that of general relativity, but we chose not to include it because of our poor knowledge of $J_2$. Both effects simply add a prograde contribution to the precession (see equation \ref{eq:intsrf}), and the absence of the $J_2$ term does not affect the overall conclusions of the paper.

\section{Precession rate and growth rate}
\label{sec:amdint}

\subsection{Angular momentum deficit}
\label{sec:amd}

By analogy with celestial mechanics, it is useful to consider the angular momentum deficit (AMD) of the system. For small eccentricities, the total AMD of the disc is
\begin{equation}
A_\mathrm{d} = \int\frac{1}{2}\ee\Sigma r^2\Omega\,2\pi r \,\id r.
\end{equation}
In this Section and in Section \ref{sec:int}, all integrals are carried from $\rin$ to $\rout$, the radial extent of the disc. With appropriate boundary conditions, the AMD is conserved for adiabatic discs, in the absence of viscous damping or resonances \citep[see][and details in Appendix \ref{app:discret}]{go06}.

A planet with a complex eccentricity $\Ep$ has an associated AMD of the form (for small $|\Ep|$)
\begin{equation}
A_{\rm p} =\frac{1}{2}|\Ep|^2\Mp\ap^2\Op.
\end{equation}
When secular interactions between the disc and the planet are included (equations \ref{eq:epd1} and \ref{eq:epd2}), AMD is exchanged between the disc and the planet. For an adiabatic, inviscid disc, the conserved quantity is now the total AMD of the system,
\begin{equation}
A = A_{\rm d} + A_{\rm p}.
\end{equation}
Including the self-gravity of the disc (equation \ref{eq:esg}) does not affect the AMD conservation. However, resonances will affect the AMD. When resonances are included (equations \ref{eq:elrd}--\ref{eq:elrp} and \ref{eq:ecrd}--\ref{eq:ecrp}), the rate of change of AMD associated with each ELR and ECR is
\begin{align}
\left(\frac{\id A}{\id t}\right)_{\rm ELR} &= \int \frac{G\Mp^2}{M_*}\Sigma |\mathscr{A}E - \mathscr{B}\Ep|^2\nonumber\\
&\times w_{\rm L}^{-1}\Delta\left( \frac{r-\rres}{w_{\rm L}} \pm 1 \right) 2\pi r \,\id r,
\end{align}
\begin{align}
\left(\frac{\id A}{\id t}\right)_{\rm ECR} & = \int \pm \dd{\ln (\Sigma/\Omega)}{\ln r}\frac{G\Mp^2}{M_*}  \Sigma |\mathscr{C}E - \mathscr{D}\Ep|^2\nonumber\\
&\times w_{\rm C}^{-1}\Delta\left( \frac{r-\rres}{w_{\rm C}} \right) 2\pi r \,\id r.
\end{align}
We note that ELRs always act to increase the AMD, with an effect that depends on the linear combination $\mathscr{A}E-\mathscr{B}\Ep$ of the complex eccentricities of the planet and the disc near the resonant location (and is approximately proportional to $E-\Ep$). Individual ECRs can either increase or decrease the AMD depending on their location and the local vortensity gradient. ECR saturation, albeit a non-linear effect, should be considered here as it can operate at very small eccentricities. Linear equations can be obtained in the two bracketing cases in which ECRs are either completely unsaturated or completely saturated.

Viscosity acts to damp the AMD at a rate 
\begin{equation}
\left(\frac{\id A}{\id t}\right)_{\rm visc} = -\int\frac{1}{2}\alpha Pr^2 \left|\pd{E}{r} \right|^2 2\pi r \,\id r.
\end{equation} 
In order to cause eccentricity growth, the net effect of mean-motion resonances on the AMD should exceed that of viscous damping.

\subsection{Normal modes and integral relations}
\label{sec:int}

The simplest solutions of our system are normal modes of the form $E(r)\,\me^{\im\omega t}$, where $\omega$ is a complex eigenfrequency.  Such a mode corresponds to a fixed distribution of elliptical orbits of the disc and planet(s), which precesses at a rate given by the real part of $\omega$ and grows at a rate given by minus the imgainary part of $\omega$.  If the system supports one or more normal mode with a positive growth rate, we may expect the fastest-growing mode to dominate the evolution of the system until it grows into a nonlinear regime.

Using the integral relations we derived in Appendix \ref{app:discret}, and adding contributions from the planet, self-gravity and short-range forces, we find that the precession rate can be written
\begin{equation}
\Re(\omega) = \frac{I_{\rm p 1} + I_{\rm p 2} + I_{\rm na} + I_{\rm 3D} + I_{\rm dd} + I_{\rm pd} + I_{\rm SRF}}{2A}.
\end{equation}
The different terms are defined as follows:
\begin{equation}
\label{eq:intp1}
I_{\rm p1} = -\int \frac{1}{2}\gamma_1 Pr^2 \left|\pd{E}{r}\right|^2 2\pi r \,\id r,
\end{equation}
with $\gamma_1=\gamma$ for the 2D adiabatic model, $\gamma_1=2-\gamma^{-1}$ for the 3D adiabatic model, and $\gamma_1=1$ for the locally isothermal model;
\begin{equation}
\label{eq:intp2}
I_{\rm p2} = \int \frac{1}{2}\gamma_2 \dd{P}{r} r \left|E\right|^2 2\pi r \,\id r,
\end{equation}
with $\gamma_2=4-3\gamma^{-1}$ for the 3D adiabatic model, and $\gamma_2=1$ otherwise;
\begin{equation}
\label{eq:intna}
I_{\rm na} = \int \frac{1}{2}\Sigma\dd{\cs}{r}r^2 e\pd{e}{r}\,2\pi \,\id r,
\end{equation}
a non-adiabatic contribution for the locally isothermal model only, and $I_{\rm na}=0$ for the adiabatic model;
\begin{align}
\label{eq:int3d}
I_{\rm 3D} &= \int \frac{3}{2}\left(1+\frac{1}{\gamma}\right)P\left|E\right|^2 2\pi r \,\id r\nonumber\\
&\hbox{or} \int \frac{3}{2r}\Sigma\dd{}{r}(\cs r^2) \left|E\right|^2 2\pi r \,\id r
\end{align}
for 3D adiabatic or locally isothermal models respectively;
\begin{align}
\label{eq:intdd}
I_{\rm dd} & = \iint \frac{1}{4}G\Sigma(r)\Sigma (r') \bigg\{ \left[ K_1 (r,r')+K_2 (r,r') \right]|E(r)-E(r')|^2\nonumber\\
& + \left[ K_1 (r,r')-K_2 (r,r') \right]|E(r)+E(r')|^2\bigg\} 2\pi r \,\id r \, 2\pi r' \,\id r'
\end{align}
for self-gravity;
\begin{align}
\label{eq:intpd}
I_{\rm pd} & = \int \frac{1}{2}G\Mp\Sigma (r) \bigg\{ \left[ K_1 (r,\ap)+K_2 (r,\ap) \right]|E(r)-\Ep|^2\nonumber\\
& + \left[ K_1 (r,\ap)-K_2 (r,\ap) \right]|E(r)+\Ep|^2\bigg\} 2\pi r \,\id r,
\end{align}
for secular disc--planet interaction, and
\begin{equation}
\label{eq:intsrf}
I_{\rm SRF} = \int f(r)\Sigma |E|^2 2\pi r \,\id r + f(\ap) \Mp |\Ep|^2
\end{equation}
for short-range forces, where
\begin{equation}
  f(r)=\frac{3G^2M_*^2}{c^2r^2}+\frac{3J_2GM_*R_*^2}{2r^3}.
\end{equation}
It can be shown that $K_1>K_2$, and therefore $I_{\rm dd}$ and $I_{\rm pd}$ are positive definite. Gravitational coupling leads to prograde precession, as do the short-range forces. In the 2D case, which was already studied by \citet{go06}, we see that the two terms associated with pressure cause a retrograde precession of the mode, provided that the pressure gradient is negative. The 3D term gives rise to a prograde precession rate, and so do the terms coming from the planet and short-range forces. The total precession rate of the mode depends on the net contribution of all these effects.

Similarly, the growth rate reads
\begin{equation}
\label{eq:intstot}
-\Im(\omega) = \frac{S_{\rm ELR} + S_{\rm ECR} + J_{\rm na} + J_{\rm visc}}{2 A},
\end{equation}
where
\begin{align}
\label{eq:selr}
S_{\rm ELR} & = \sum_{\rm ELR}\int \frac{G\Mp^2}{M_*}\Sigma |\mathscr{A}E - \mathscr{B}\Ep|^2\nonumber\\
&\times w_{\rm L}^{-1}\Delta\left( \frac{r-\rres}{w_{\rm L}} \pm 1 \right) 2\pi r \, \id r,
\end{align}
\begin{align}
\label{eq:secr}
S_{\rm ECR} & = \sum_{\rm ECR}\int\pm \dd{\ln (\Sigma/\Omega)}{\ln r}\frac{G\Mp^2}{M_*}  \Sigma |\mathscr{C}E - \mathscr{D}\Ep|^2\nonumber\\
&\times w_{\rm C}^{-1}\Delta\left( \frac{r-\rres}{w_{\rm C}}\right) 2\pi r \,\id r,
\end{align}
\begin{equation}
\label{eq:jvisc}
J_{\rm visc} = -\int \frac{1}{2}\ab Pr^2\left|\pd{E}{r}\right|^2 2\pi r \, \id r
\end{equation}
and
\begin{equation}
\label{eq:jna}
J_{\rm na} = \int \frac{1}{2}\Sigma \dd{\cs}{r}r^2 e^2 \pd{\varpi}{r}\,2\pi r \,\id r,
\end{equation}
for eccentric Lindblad resonances, eccentric corotation resonances, viscosity and non-adiabatic effects (in the locally isothermal model only), respectively. It clearly shows that the contribution from each ELR always acts to increase the growth rate, while viscosity reduces it. The effect of each ECR depends on the local vortensity gradient, but their net effect usually contributes to the damping of eccentricity. It is important to notice that, in expression (\ref{eq:intstot}) for the growth rate, the AMD of the mode appears in the denominator. In anticipation of the solutions in terms of normal modes that will be presented in the remainder of the paper, we can expect that, if an eccentric mode is confined in the inner parts of the disc, and decays towards zero in the outer parts, it should present a higher growth rate than a less confined mode, if the source terms due to resonances are comparable.

\section{Solution in terms of normal modes}
\label{sec:solmode}

\subsection{Numerical methods}

In order to solve the eccentricity equations of Section \ref{sec:evo_eqs}, we seek normal modes of the form $E(r)\,\me^{\im\omega t}$ for the complex eccentricity. We can then solve the resulting eigenvalue problem for the (usually complex) frequencies $\omega$. This is achieved by dividing the disc into $N$ rings, spaced equally in the variable $\ln r$, and associating an eccentricity to each ring. The discrete model that results from this scheme is described in Appendix \ref{app:discret}. We use LAPACK to solve for the eigenvalues and eigenvectors of the generalized eigenvalue problem. Some spurious solutions can be generated that show oscillations on the grid scale, and which should be discarded. Each mode represents a radial eccentricity distribution, and its precession and growth rates are given by the real and (negative) imaginary part of the eigenfrequency, respectively.
As a measure of the mode order, we take the number of zeros of the real part of the eigenvector. While this is strictly valid only for a ``classical'' Sturm-Liouville problem with real functions of $r$, we also use it as an indicator of the mode order for our more general problem. When necessary, integrals are solved numerically using a Romberg method \citep{press92}.

The disc extends from an inner radius $\rin$ to an outer radius $\rout$. For a disc with free boundaries at both edges, the boundary conditions are 
\begin{equation}
\ff\pd{E}{r} = 0
\end{equation}
at $\rin$ and $\rout$, where $\ff$ is a function of $r$ defined in Appendix~\ref{app:discret}. These conditions are automatically satisfied by our prescription for the surface density and pressure, as both fall to zero at each edge (see below).

The grid resolution has to be such as the low-order modes that are of interest here do not show any variation with resolution. We find that dividing the disc into $N=1000$ rings gives consistent results. A good internal check on the accuracy can be obtained by comparing the real and imaginary parts of the eigenfrequency with the ones obtained using the integral formulae in Section \ref{sec:int}. 

\subsection{Units}

We choose to work in physical units (astronomical units, solar masses and years). For most parts, the disc--planet interactions could be described in arbitrary units, but the inclusion of short-range forces such as the general relativistic correction, which depend on the distance to the star, make it more convenient to choose physical units. Unless mentioned otherwise, the precession and growth rates are given in units of the planet's orbital frequency. We keep this convention even when the planet is not included, in order to help the comparison between the various models. In our fiducial example, the planet's orbital frequency is calculated at $\ap=0.049~\text{au}~(\sim 4~{\rm d})$, a distance typical of hot Jupiters. Finally we recall that since we are solving a normal mode problem, the solution can be scaled by any factor and should not be interpreted as the true eccentricity, but as being arbitrarily normalized. 
 
\subsection{Disc model}
\label{sec:discmodel}

In order to solve the eccentricity equation, one needs a description of the disc structure. Following \citet{bl94}, we note that in the inner parts of the disc (from $\sim 0.1$ to $\lesssim 10~\text{au}$), the temperature ranges from 100 to 1000 K. In this region the opacity is mostly proportional to $T^{1/2}$. This allows us to construct a simple disc model in which the surface density is given by
\begin{equation}
\Sigma = \Sigma_0\left(\frac{r}{r_0}\right)^{-1/2} \left(1 - \sqrt{\frac{\rin}{r}}\right)^{5/9}\tanh\left(\frac{\rout-r}{\wout}\right),
\end{equation}
and the aspect ratio follows
\begin{equation}
\label{eq:hoverr}
\frac{H}{r} = h_0\left(1-\sqrt{\frac{\rin}{r}}\right)^{2/9},
\end{equation}
where $\Sigma_0$ and $r_0$ are characteristic values of surface density and radius, and $h_0$ is the limiting value of the aspect ratio at large radius.  $H$ is defined such that the vertically integrated pressure is
\begin{equation}
P = \Sigma \Omega^2 H^2,
\end{equation}
as follows from vertical hydrostatic equilibrium. With our choice for $\Sigma$ and $H/r$, we have
\begin{equation}
P = P_0\left(\frac{r}{r_0}\right)^{-3/2}\left(1-\sqrt{\frac{\rin}{r}}\right)  \tanh\left(\frac{\rout-r}{\wout}\right).
\end{equation}
In the locally isothermal case, we relate the isothermal sound speed to $H$ through
\begin{equation}
c_{\rm s} = H \Omega.
\end{equation} 

\begin{figure*}
    \begin{center}
    \includegraphics[width=2\columnwidth]{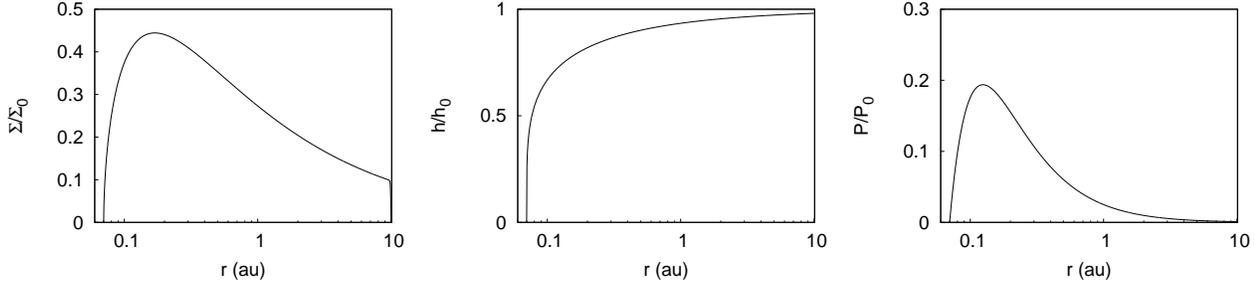}
    \caption{Normalized surface density, aspect ratio and pressure as functions of radius (in logarithmic scale).}
    \label{fig:psh}
    \end{center}
\end{figure*}

Apart from the $\tanh$ factor, this model corresponds to a steady accretion disc with constant alpha viscosity and opacity $\propto T^{1/2}$. In this model, the surface density, aspect ratio and pressure naturally go to zero at the inner edge of the disc. At the outer edge, we also wish to have zero surface density and pressure, which is achieved by using a $\tanh$ taper function, in which the drop to zero occurs on a width defined through $\wout$. Here $\wout$ is taken to be $0.01\rout$. The value has to be chosen according to the resolution so that quantities smoothly drop off to zero over a few grid points. The exact value has no important effect on the solution. We plot the surface density, aspect ratio and pressure radial profiles in Fig.~\ref{fig:psh}.

In this model, one expects the surface density to be around $1000~\text{g cm}^{-2}$ at $r_0=0.1~\text{au}$ \citep{bckh97}. This implies that $\Sigma_0$ is of the order a few times $10^{-4}$ in our units.

\section{A non-self-gravitating and inviscid disc}
\label{sec:fidu}

In this section we solve the normal mode problem described in Section \ref{sec:solmode}, where the eccentricity is assumed to be of the form $E(r)\,\me^{\im \omega t}$. We start with the case of a disc in which viscosity, self-gravity and resonant interactions with the planet are excluded. However, we keep the secular part of the planet--disc interaction. The eccentricity evolution is governed by one of the models described by equations (\ref{eq:2dadia}) to (\ref{eq:3diso}), and equation (\ref{eq:epd1}) with $\Ep=0$.

We first illustrate the solution for a disc that extends from $\rin=0.07$ au to $\rout=10$ au, with a mass ratio $q_\mathrm{d}=0.005$, an aspect ratio $h_0=0.05$, and a viscosity $\ab=4\times10^{-3}$. In the adiabatic case, we take the adiabatic index to be  $\gamma=1.4$\footnote{\citet{bbk13} pointed out that the adiabatic index can vary from 5/3 to 7/5 with increasing disc temperature. However the exact value of the adiabatic index does not significantly change our results.}. The planet is on a fixed circular orbit at $\ap=0.049$ au, with a mass ratio  of $q_\mathrm{p}=0.003$, and we neglect the self-gravity of the disc.

The total mass of the disc is a somewhat arbitrary parameter, since for a given surface density profile it depends on the inner and outer disc radius. Here our mass is chosen such as $\Sigma_0=1.3\times 10^{-4}$, in the range of values expected for such discs. It ensures that self-gravity can be neglected. We note here that if the disc--planet interaction is localized in the inner part of the disc, the total disc mass is irrelevant, as most of it is located in the outer part. A more relevant quantity is the local disc mass near the inner edge which we define as $\pi\Sigma_0\rin^2$. With our disc parameters, the local disc mass is expected to be much lower than the mass of a Jovian planet.

\subsection{Bound states in a potential well}
\label{sec:schrod}
In the absence of self-gravity and viscosity, the eccentricity propagates through the disc solely by means of pressure, as a one-armed density wave.  The evolutionary equations of Section~\ref{sec:pres} are dispersive wave equations related to the Schr\"{o}dinger equation of quantum mechanics.  
A basic understanding of the physics at play can be achieved by rewriting the equation for an eccentric normal mode as a time-independent Schr\"{o}dinger equation:
\begin{equation}
-\frac{\id^2 \psi}{\id x^2} + \left[V(x)-\mathcal{E}\right]\psi = 0,
\end{equation} 
where $x$ is a new independent variable related to $r$, $\psi(x)$ is the wavefunction, $V(x)$ is an effective potential and $\mathcal{E}$ is the effective energy eigenvalue.  In the following we assume that the sound speed can be written $c_{\rm s}=\epsilon r \Omega$, where $\epsilon$ is taken to be the characteristic aspect ratio $h_0$. In Appendix \ref{app:schrod} we present a derivation in which $H/r$ can take a more general form such as the one we use in Equation (\ref{eq:hoverr}). The simple model that we use here is sufficient to capture the relevant physical mechanism.

In the case of a locally isothermal disc, the transformation required to obtain a Schr\"{o}dinger equation is
\begin{align}
\label{eq:schrodiso}
\psi &\propto r^{11/8}\Sigma^{1/2}E, \nonumber\\
x&=\left(\frac{r}{\ap}\right)^{3/4},\nonumber\\
\mathcal{E} &= -\frac{32}{9\epsilon^2}\frac{\omega}{\Op}, \nonumber\\
V &= \frac{1}{f} \frac{\id^2 f}{\id x^2} - \frac{32}{9x^2}\dd{\ln \Sigma}{\ln r} - \frac{32}{9\epsilon^2} \frac{\qp}{x^{2/3}}\ap K_1(r,\ap),
\end{align}
where $f=r^{11/8}\Sigma^{1/2}$. In the 3D case, an additional term $-16/(3x^2)$ is added to $V$.

In the adiabatic case we have
\begin{align}
\label{eq:schrodadia}
\psi &\propto r^{11/8}P^{1/2}E, \nonumber\\
x&=\left(\frac{r}{\ap}\right)^{3/4}, \nonumber\\
\mathcal{E} &= -\frac{32}{9a\epsilon^2}\frac{\omega}{\Op}, \nonumber\\
V &= \frac{1}{f} \frac{\id^2 f}{\id x^2} - \frac{16}{9x^2}\frac{b}{a}\dd{\ln P}{\ln r} - \frac{16}{9x^2}\frac{c}{a} - \frac{32}{9a\epsilon^2} \frac{\qp}{x^{2/3}}\ap K_1(r,\ap),\nonumber\\
\end{align}
where $f=r^{11/8}P^{1/2}$. In 2D, we have $a=\gamma$, $b=1$ and $c=0$, while in 3D we have $a=2-\gamma^{-1}$, $b=4-3\gamma^{-1}$, $c=3(1+\gamma^{-1})$.

\begin{figure*}
    \begin{center}
    \includegraphics[width=2\columnwidth]{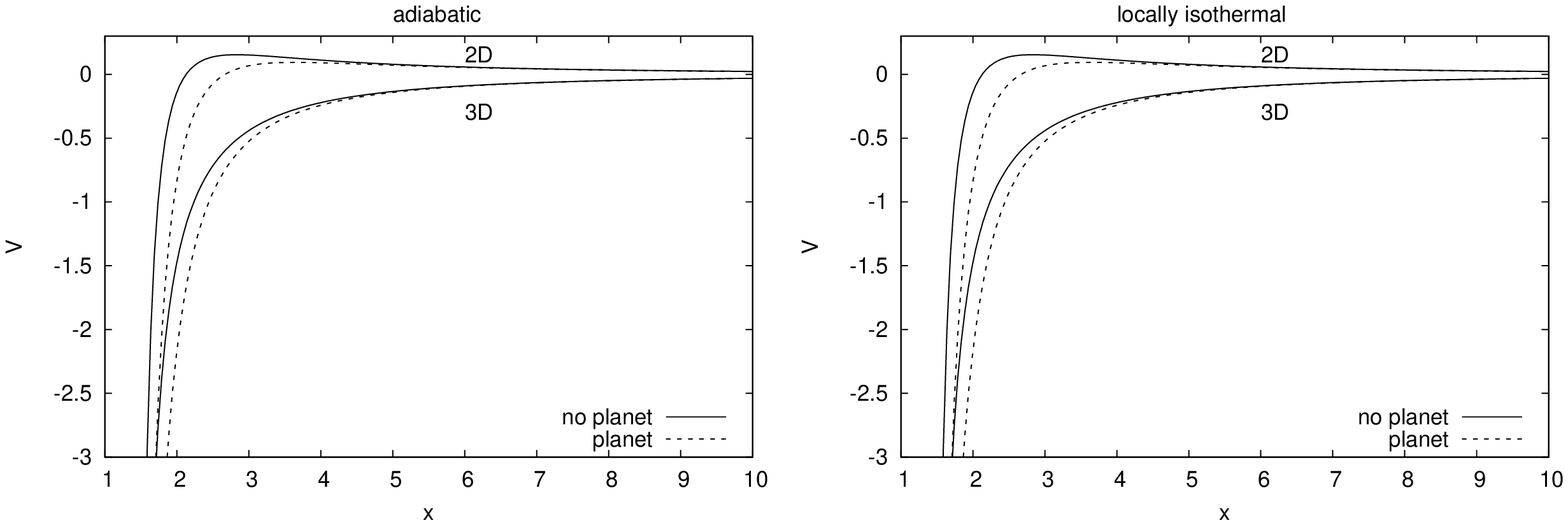}
    \caption{Effective potential for the analogous Schr\"odinger equation, for the 2D and 3D adiabatic (left panel) and locally isothermal (right panel) models. The solid line is the potential without the planet, while the dotted line includes the planet's contribution. The planet has $\ap=0.049$, $\qp=0.003$, and the disc has $\epsilon=0.05$.}
    \label{fig:pot}
    \end{center}
\end{figure*}

In both cases, the shape of the effective potential $V$ can help in understanding the trapping of the modes by analogy with bound states of the Schr\"{o}dinger equation. Typical potentials are plotted in Fig.~\ref{fig:pot}. The results are rather similar in the locally isothermal and adiabatic cases. The profiles show that a deep potential well exists in 3D, which could support a bound state in the form of a confined eccentric mode that decays at $r\to\infty$. Since $V\to 0$ as $r\to\infty$, a bound state has $\mathcal{E}<0$, and is therefore a prograde mode $(\omega>0)$. In 2D, the potential well is less deep, and bound states might not exist. This distinction, which arises from the additional prograde precession due to pressure in a 3D disc, was identified as important by \citet{ogilvie08}. The potential arising from the planet also contributes to deepening the potential well. It will turn out to be of importance for helping the confinement of 2D growing modes.

\subsection{Mode analysis for a non-self-gravitating disc}
\label{sec:fiduP}

We start by solving the eigenvalue problem for a simple case of a non-self-gravitating, inviscid disc without a planet, short-range forces or self-gravity. We use the parameters given at the beginning of Section \ref{sec:fidu}.

\begin{figure*}
    \begin{center}
    \includegraphics[width=2\columnwidth]{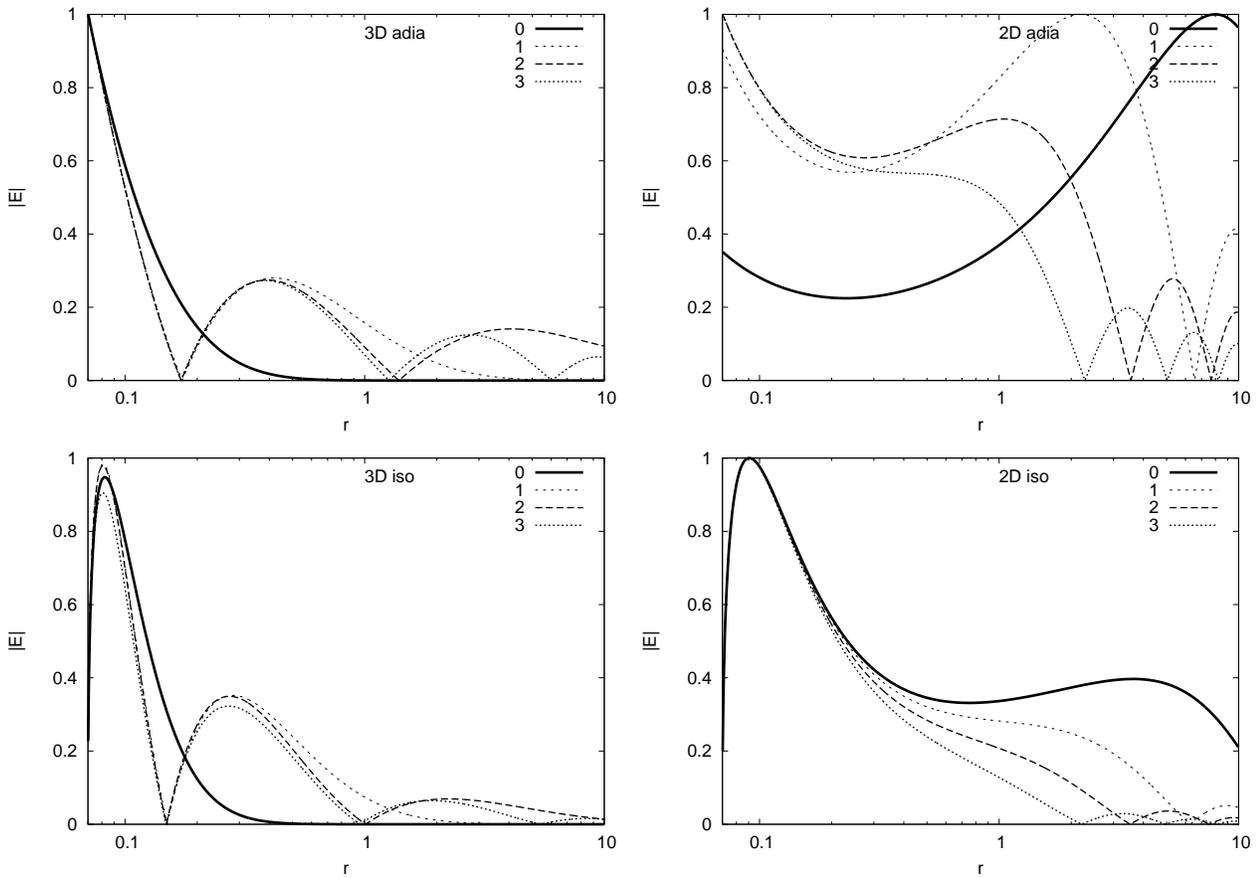}
    \caption{The four lowest-order modes of our fiducial example for an inviscid, non-self-gravitating disc without a planet, for the four different models: 3D adiabatic, 2D adiabatic, 3D isothermal and 2D isothermal. The number in the top right corner indicates the order of the corresponding mode. Their precession rates are given in Table \ref{tab:fiduP}.}
    \label{fig:fiduP}
    \end{center}
\end{figure*}

In Fig.~\ref{fig:fiduP}, we represent the mode shape of the four lowest-order modes. In this simple model, major differences between 2D and 3D adiabatic and isothermal discs appear. In 3D, there exists a mode of order 0 which is confined in the inner parts of the disc, in agreement with what we speculated in Section \ref{sec:schrod}. This mode decays on the characteristic length of 1 au, and is insensitive to the outer boundary condition as long as $\rout\gtrsim 1~\text{au}$. Higher-order modes do not always decay to zero at large radii, and might therefore be sensitive to the outer radius. However they contain a significantly larger amount of AMD, and are therefore expected to grow less rapidly, as we will confirm in Section \ref{sec:fiduvr}. In Table \ref{tab:fiduP} we show the precession rates of these modes. The disc supports three prograde modes, all the others being retrograde. Although we do not include a planet in this calculation, we still give the precession and growth rates in units of the planet's orbital frequency at $\ap=0.049$ au, since it will help comparing with the next section where we include the planet.

In 2D, there no longer exists a mode which is trapped in the inner parts of the disc (see right panels of Fig.~\ref{fig:abgrt}). This is especially true in the adiabatic case where all modes show a large eccentricity distribution at large radii. As can be seen in Table \ref{tab:fiduP}, 2D discs only support a set of retrograde modes. 

\begin{table*}
  \begin{tabular}{crrrr}   
  \multicolumn{1}{c}{Mode order} & \multicolumn{1}{c}{3D adiabatic} & \multicolumn{1}{c}{2D adiabatic} & \multicolumn{1}{c}{3D isothermal} & \multicolumn{1}{c}{2D isothermal} \\ \hline 
0 & $ 5.727\times 10^{-4} $ & $-1.863\times 10^{-6} $ & $ 7.615\times 10^{-4} $ & $-2.499\times 10^{-6} $ \\
1 & $ 2.250\times 10^{-5} $ & $-1.088\times 10^{-5} $ & $ 2.833\times 10^{-5} $ & $-9.179\times 10^{-6} $  \\
2 & $ 1.059\times 10^{-6} $ & $-2.553\times 10^{-5} $ & $ 7.289\times 10^{-7} $ & $-1.968\times 10^{-5} $ \\
3 & $-6.579\times 10^{-6} $ & $-4.574\times 10^{-5} $ & $-4.727\times 10^{-6} $ & $-3.346\times 10^{-5} $  \\
	\hline
  \end{tabular}
  \caption{Precession rate of the four lowest-order modes of our fiducial example, for an inviscid, non-self-gravitating disc without planets, in the 2D and 3D abiabatic and isothermal cases. The shape of the corresponding modes is shown in Fig.~\ref{fig:fiduP}.}
  \label{tab:fiduP}
\end{table*}

\subsection{Secular perturbations from a planet}
\label{sec:fidusec}

The secular perturbation from the planet gives rise to an additional source of prograde precession in the disc. Previously, we have noted that the planet helps to deepen the effective potential well, and could therefore assist the trapping of a mode in the inner part of the disc (see Section \ref{sec:schrod} and Fig.~\ref{fig:pot}). 

Our choice of semi-major axis is such that $\ap/\rin=0.7$. The planet is expected to migrate into the cavity via tidal interaction with the disc, until its 1:2 outer Lindblad resonance (OLR) lies just inside the inner edge of the disc, which here would be $\ap/\rin\approx 0.63$. The planet should then remain at a constant semi-major axis, a convenient setup in our secular framework. With our choice of surface density profile, motivated by the truncation of the disc by a magnetospheric cavity, the surface density falls to zero at $r=\rin$ and exerts no torque on the planet within this radius. Instead we expect the planet to mostly interact with the peak of the surface density just outside the inner edge. Therefore the planet should stop slightly closer to the disc than the 1:2 orbital commensurability with the inner edge, and we choose the fiducial value of $\ap/\rin=0.7$. We point out that in the study of planets in cavities by \citet{rice08}, Jupiter-mass planets stalled at orbital distances $\ap/\rin=0.6~\text{to}~0.7$ over the few hundred orbits over which the simulations were performed. We recall that in our secular treatment of disc--planet interactions, it was assumed that the semi-major axis of the planet was constant in time. We discuss the influence of $\ap$ in Section \ref{sec:paramap}.

\begin{figure*}
    \begin{center}
    \includegraphics[width=2\columnwidth]{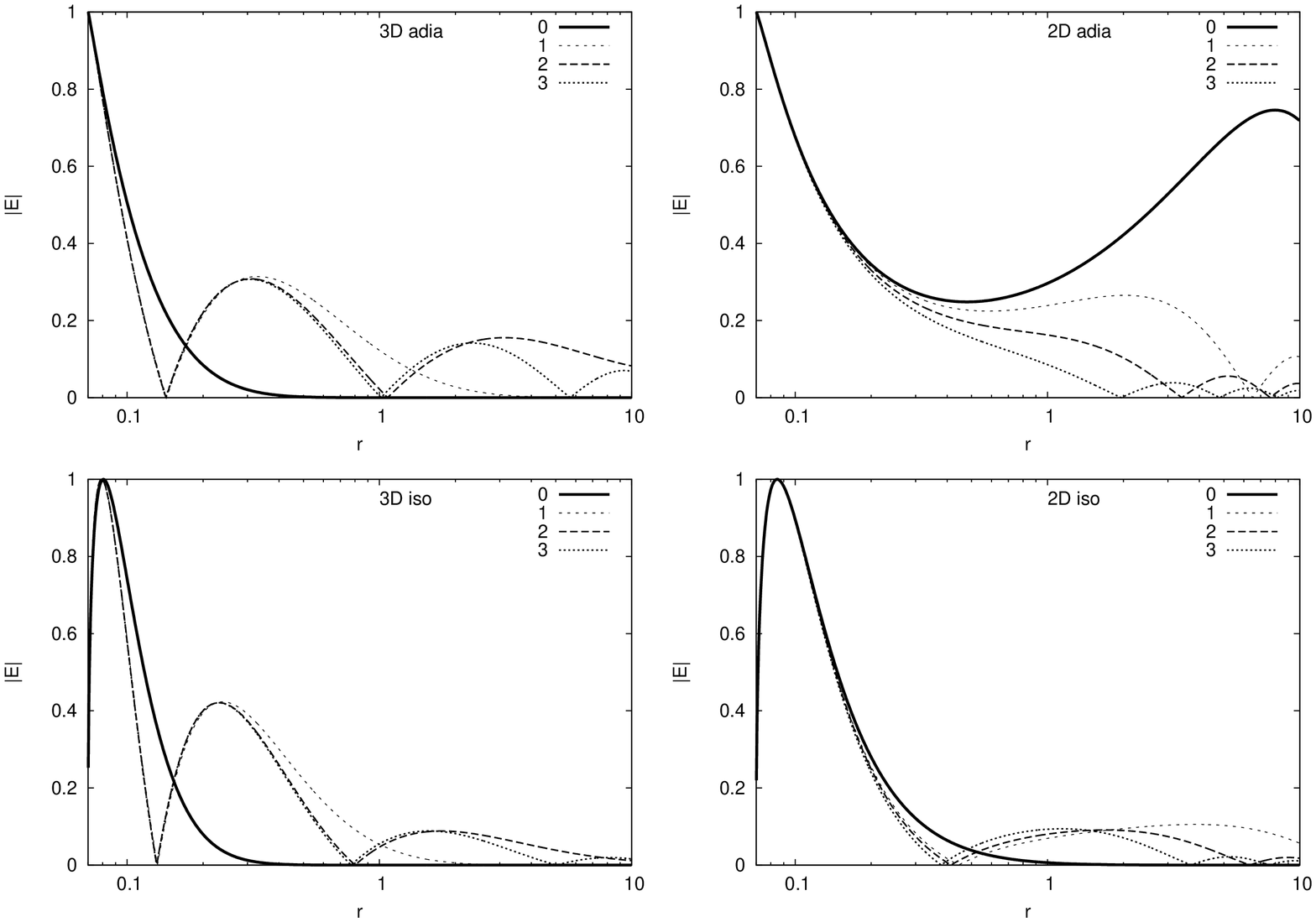}
    \caption{The four lowest-order modes of our fiducial example for an inviscid non-self-gravitating disc secularly perturbed by a planet (but with no resonances or back-reaction on the planet), for the four different models: 3D adiabatic, 2D adiabatic, 3D isothermal and 2D isothermal. The number in the top right corner indicates the order of the corresponding mode. Their precession rates are given in Table \ref{tab:fiduPP}.}
    \label{fig:fiduPP}
    \end{center}
\end{figure*}

In Fig.~\ref{fig:fiduPP} we show the mode shape, and the associated precession rates are displayed in Table \ref{tab:fiduPP}. In 3D, the effect of adding a planet on the mode shape is negligible, as the mode is already trapped in the deep effective potential due to 3D pressure effects and discussed in Section \ref{sec:schrod}. The planet enhances the prograde precession of these modes, but does not strongly affect the retrograde modes, which are pure disc modes. In 2D, the planet is observed to reduce the amplitude of the mode in the outer parts of the disc compared to that of the inner parts. Here a significant difference arises between the 2D isothermal and adiabatic cases. In the isothermal case, the additional contribution of the planet to the potential well is enough to confine the lowest-order mode in the inner parts of the disc, while in the adiabatic case this mode `leaks' into the outer disc. When resonant and viscous effects are added in the next section, this will have important consequences for the growth of the mode. In Table \ref{tab:fiduPP} we can see that the lowest-order 2D isothermal mode has become prograde owing to the inclusion of the planet, while the 2D adiabatic modes are only weakly affected by the planet (see Table \ref{tab:fiduP} for comparison). In the example that we show here, the perturbation from the planet is not strong enough to cause any mode to precess progradely in the 2D adiabatic calculation. However more massive planets closer to the disc's inner edge could assist the confinement of a 2D progade mode. 

\begin{table*}
  \begin{tabular}{crrrr}   
  \multicolumn{1}{c}{Mode order} & \multicolumn{1}{c}{3D adiabatic} & \multicolumn{1}{c}{2D adiabatic} & \multicolumn{1}{c}{3D isothermal} & \multicolumn{1}{c}{2D isothermal} \\ \hline 
0 & $ 9.230\times 10^{-4} $ & $-1.856\times 10^{-6} $ & $ 1.121\times 10^{-3} $ & $ 2.559\times 10^{-5} $ \\
1 & $ 3.298\times 10^{-5} $ & $-1.069\times 10^{-5} $ & $ 4.099\times 10^{-5} $ & $-2.518\times 10^{-6} $  \\
2 & $ 1.509\times 10^{-6} $ & $-2.432\times 10^{-5} $ & $ 1.156\times 10^{-6} $ & $-9.462\times 10^{-6} $ \\
3 & $-5.215\times 10^{-6} $ & $-4.138\times 10^{-5} $ & $-3.849\times 10^{-6} $ & $-2.097\times 10^{-5} $  \\
	\hline
  \end{tabular}
  \caption{Precession rate of the four lowest-order modes of our fiducial example, for an inviscid, non-self-gravitating disc, with the secular perturbation from a planet ($\qp=0.005, \ap=0.07$), in the 2D and 3D abiabatic and isothermal cases. The shape of the corresponding modes is shown in Fig.~\ref{fig:fiduPP}.}
  \label{tab:fiduPP}
\end{table*}

\section{A more realistic example}
\label{sec:fiduall}

In this Section, we progressively add more physical effects to our disc model, and show how they affect the shape, precession rate and growth rate of the modes. The physical parameters for the disc and the planet are the same as in Section \ref{sec:fidu}. 

\subsection{Viscosity and resonant interactions}
\label{sec:fiduvr}

\begin{figure*}
    \begin{center}
    \includegraphics[width=2\columnwidth]{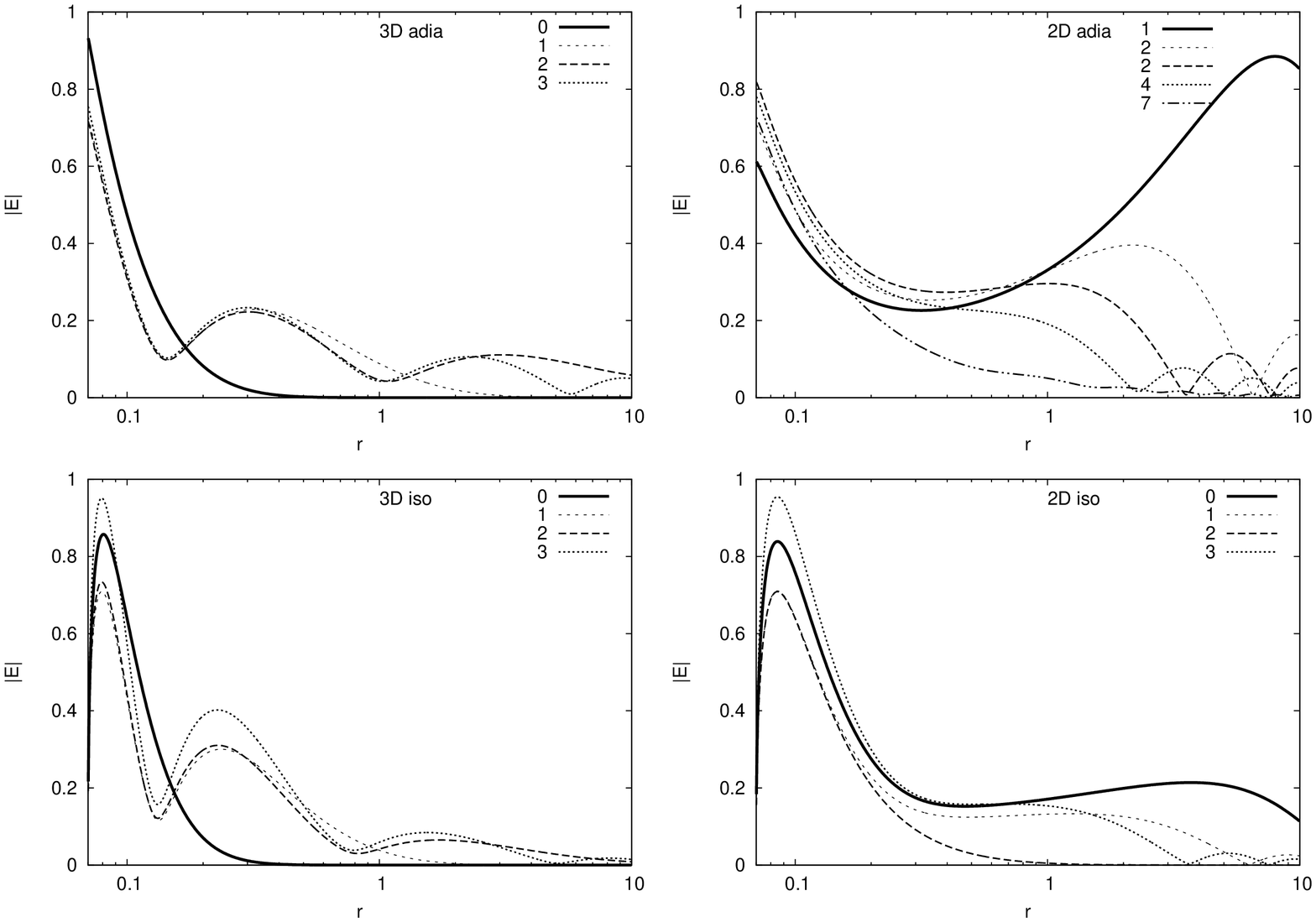}
    \caption{The four lowest-order modes of our fiducial example for a viscous, non-self-gravitating disc, including secular and resonant interactions with a planet, for the four different models: 3D adiabatic, 2D adiabatic, 3D isothermal and 2D isothermal. The number in the top right corner indicates the order of the corresponding mode. Their precession rates are given in Table \ref{tab:fiduPPVR}.}
    \label{fig:fiduPPVR}
    \end{center}
\end{figure*}

We first add effects that will contribute to the growth rate: viscosity, and eccentric Lindblad and corotation resonances. The planet is kept on a fixed circular orbit. The shape of the modes is shown in Fig.~\ref{fig:fiduPPVR}, and the corresponding precession rates and growth rates are displayed in Table \ref{tab:fiduPPVR}. The shape of the 3D modes is slightly different from Fig.~\ref{fig:fiduPP} because the eccentricity now has an imaginary component that shifts the location of local minima away from zero. However, comparing Table \ref{tab:fiduPPVR} with Table \ref{tab:fiduPP} shows that the precession rate of these modes remain similar. In 3D, the lowest-order mode, which is confined in the inner parts of the disc, contains a very small amount of AMD. This mode therefore has the highest growth rate. The growth rates are fairly similar in the adiabatic and locally isothermal 3D cases, and the fastest growing mode can grow in $3\times 10^{3}$ orbital periods of the planet, which is short compared to the disc's lifetime if the planet has an orbital period of 3 to 4 days.

In 2D, the lowest-order modes in the adiabatic case do not present significant differences from what was found in Section \ref{sec:fidusec}. The precession rates are similar and, because no mode is confined in the inner disc, all modes have relatively low growth rates, much lower than in 3D.  However there exists a series of higher-order modes that decay slowly at large radii, but do not show the large-amplitude oscillations of the lowest-order mode. These modes are of order 6 to 10, and all show growth rates of about $10^{-5}$. We show in Fig.~\ref{fig:fiduPPVR} and Table \ref{tab:fiduPPVR} the fastest growing of these modes, which is of order 7. 

The 2D isothermal case is more subtle. As can be seen in Table \ref{tab:fiduPPVR}, the only prograde mode that was found in Table \ref{tab:fiduPP} has now disappeared. The fastest growing mode is of order 2 (see Fig.~\ref{fig:fiduPPVR}), and seems at odds with other growing modes because its growth rate is faster than its precession rate. These peculiar modes only appear because of the resonant effects, which can impose a very fast growth rate to a mode, which then decays rapidly towards the outer boundary. This decay occurs before the mode has time to propagate to the outer boundary and reflect to form a standing wave.  Because this mode is confined in the inner disc it grows faster than the fastest growing mode in the 2D adiabatic case. As we will see in Section \ref{sec:paramqp} these modes can also appear in 3D. The analogy with a Schr\"odinger equation is harder to establish in this case since the potential now contains an imaginary part as well. In particular it is possible for a mode to be trapped in a complex effective potential with a retrograde precession rate and a positive growth rate. The three other modes in the 2D adiabatic regime have an equivalent in Table \ref{tab:fiduPP}, although the mode order can be different.

\begin{table}
  \begin{tabular}{lcrr}   
  \multicolumn{1}{c}{Model} & \multicolumn{1}{c}{Mode order} & \multicolumn{1}{c}{precession rate} & \multicolumn{1}{c}{growth rate}  \\ 
\hline 
3D adiabatic &     0 & $ 8.368\times 10^{-4} $ & $ 3.469\times 10^{-4} $ \\
 & 1 & $ 3.153\times 10^{-5} $ & $ 1.030\times 10^{-5} $ \\
 & 2 & $ 1.489\times 10^{-6} $ & $ 3.889\times 10^{-7} $ \\
 & 3 & $-5.163\times 10^{-6} $ & $ 1.062\times 10^{-6} $ \\
\hline
2D adiabatic &    1 & $-1.862\times 10^{-6} $ & $ 3.061\times 10^{-9} $ \\
 & 2 & $-1.086\times 10^{-5} $ & $ 6.930\times 10^{-8} $ \\
 & 2 & $-2.549\times 10^{-5} $ & $ 4.565\times 10^{-7} $ \\
 & 4 & $-4.579\times 10^{-5} $ & $ 1.619\times 10^{-6} $ \\
 & 7 & $-1.681\times 10^{-4} $ & $ 1.978\times 10^{-5} $ \\
\hline
3D isothermal &     0 & $ 1.063\times 10^{-3} $ & $ 2.985\times 10^{-4} $ \\
 & 1 & $ 3.943\times 10^{-5} $ & $ 1.063\times 10^{-5} $ \\
 & 2 & $ 1.122\times 10^{-6} $ & $ 3.304\times 10^{-7} $ \\
 & 3 & $-3.846\times 10^{-6} $ & $ 5.880\times 10^{-7} $ \\
\hline
2D isothermal &	   0 & $-2.504\times 10^{-6} $ & $ 6.157\times 10^{-9} $ \\
 & 1 & $-9.290\times 10^{-6} $ & $ 8.612\times 10^{-8} $ \\
 & 2 & $-1.574\times 10^{-5} $ & $ 7.463\times 10^{-5} $ \\
 & 3 & $-2.042\times 10^{-5} $ & $ 4.352\times 10^{-7} $ \\
\hline
  \end{tabular}
  \caption{Precession rates and growth rates for the lowest-order modes of our fiducial example, when viscosity and secular and resonant interactions with the planet are included, in the 2D and 3D abiabatic and isothermal cases. The shape of the corresponding modes is shown in Fig.~\ref{fig:fiduPPVR}.}
  \label{tab:fiduPPVR}
\end{table}

\begin{figure}
    \begin{center}
    \includegraphics[width=1\columnwidth]{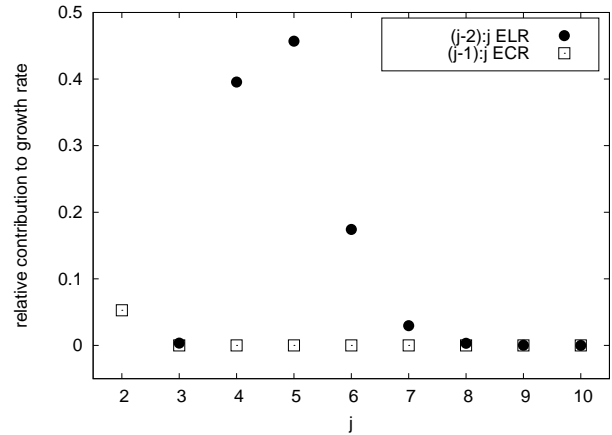}
    \caption{Contribution to the growth rate of the fastest growing mode in the 3D adiabatic model for various resonances. Eccentric Lindblad resonances (ELRs) are represented as full circles, while eccentric corotation resonances (ECRs) are empty squares. All contributions are normalized so that their sum is one. ECRs give a negative contribution to the growth rate, and we show here their absolute value. The 2:4, 3:5 and 4:6 ELRs contribute to most of the growth rate, much more than the 1:3 ELR. The only ECR to contribute significantly is the 1:2.}
    \label{fig:rsfidu}
    \end{center}
\end{figure}

Using Equations (\ref{eq:selr})--(\ref{eq:jna}), we can study the contribution of different physical effects to the growth and decay of eccentricity. In Fig.~\ref{fig:rsfidu} we show, in the case of a 3D adiabatic disc, the normalized contribution from the first few ELRs and ECRs. The contribution of ECRs is actually negative (causing a damping of eccentricity), and we plot its absolute value. The contribution from viscosity is negligible in this case. It clearly appears that the only ECR contributing to the damping is the 1:2. Regarding ELRs, the 1:3 is rather ineffective at exciting the eccentricity. The main contributions come from the 2:4 and 3:5 ELRs, with a smaller contribution from the 4:6. Although the 2:4 ELR shares the same nominal radius as the 1:2 ECR, we recall that the peak of the torque distribution of the ELR is actually slightly shifted away from the planet. In addition, the width of ELRs is significantly larger than the width of ECRs, allowing for a larger part of the disc to contribute to the growth. The consequence of the shift of the centre of the resonance is even more visible when comparing the 4:6 ELR with the 2:3 ECR. The 2:3 ECR lies in the cavity and is inoperative in the disc, while the 4:6 ELR is partly in the disc and contributes to the eccentricity growth.

\subsection{Back-reaction on the planet}
\label{sec:fiduplanet}

\begin{figure*}
    \begin{center}
    \includegraphics[width=2\columnwidth]{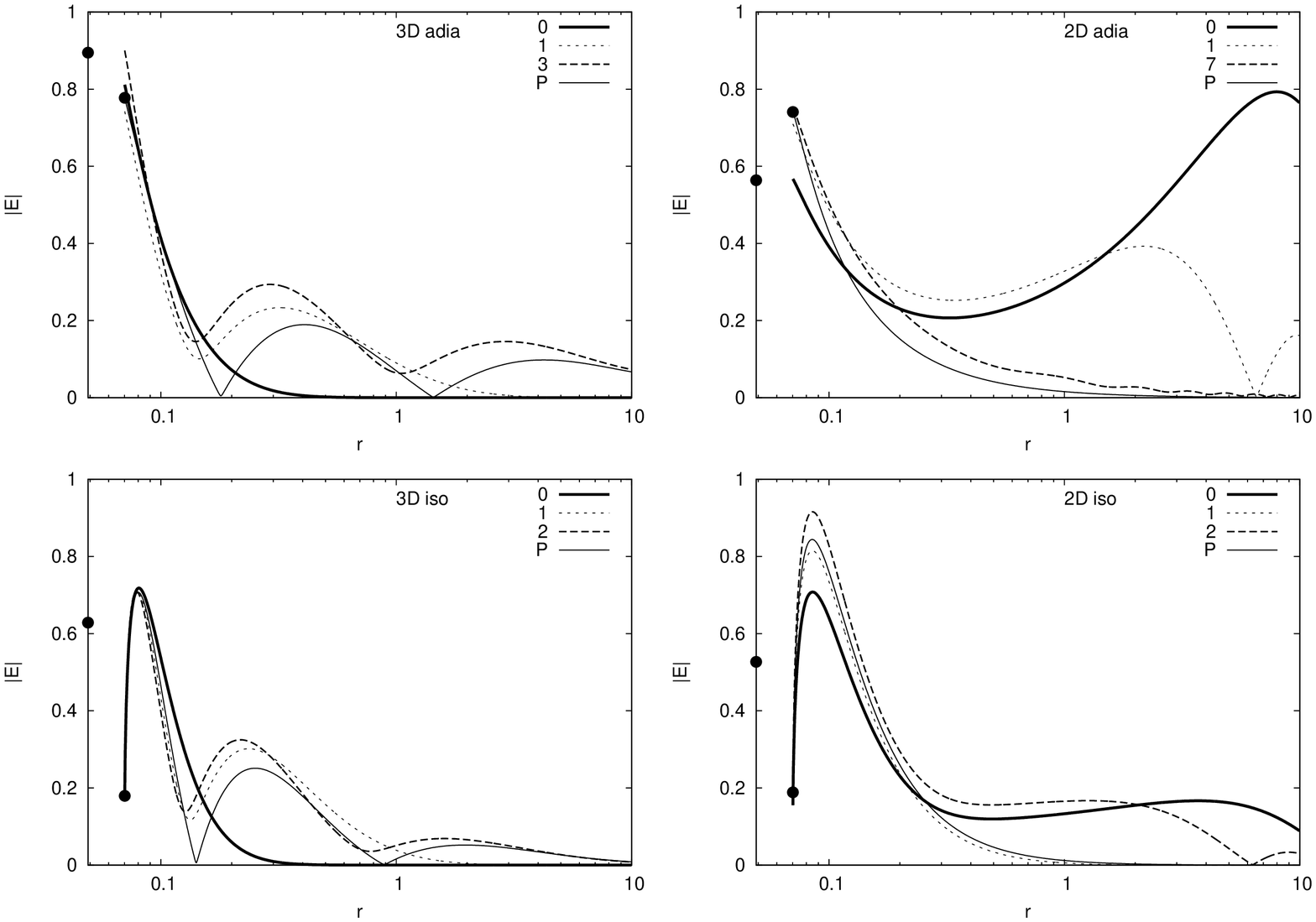}
    \caption{The four lowest-order modes of our fiducial example for a viscous, non-self-gravitating disc, including secular and resonant interactions with a planet, and back-reaction on the planet, for the four different models: 3D adiabatic, 2D adiabatic, 3D isothermal and 2D isothermal. The relativistic contribution to precession is also included. We show here the distribution of eccentricity in the disc, truncated at the disc's inner edge $\rin=0.07$. All modes but one have a negligible amount of eccentricity (or AMD) in the planet, and we do not show it here. For the mode dominated by the planet, the eccentricity in the planet is represented by a black dot at the location of the planet ($\ap=0.049$), which links to the corresponding mode in the disc. The number in the top right corner indicates the order of the corresponding mode. Their precession rates are given in Table \ref{tab:fiduPPVRPGR}.}
    \label{fig:fiduPPVRPGR}
    \end{center}
\end{figure*}

We now allow the disc to act back on the eccentricity of the planet (i.e. we do not assume $\Ep=0$ any more in all the equations where $\Ep$ appears).  Fig.~\ref{fig:fiduPPVRPGR} and Table \ref{tab:fiduPPVRPGR} show a few modes of interest for the disc, and the mode that is dominated by the planet (defined as the mode for which the AMD is predominantly in the planet, and labelled ``P''). The planet-dominated mode undergoes a small prograde precession, and has  a slow growth rate. 

We find that the growth rate of the planet-dominated mode is significantly smaller than that of the disc mode, when they both exist. Interestingly, the 2D planet mode grows on a timescale equal to, or larger than that in 3D. This effect will appear consistently for a large range of parameters described in Section \ref{sec:param}. In 2D, the planet mode does not propagate very far into the disc, and therefore its AMD is low. On the other hand, the 3D planet mode is more strongly coupled to the disc. It shows some oscillations and propagates further into the disc. Its AMD is therefore larger, reducing its growth rate. 

In this calculation, we also include the relativistic precession term. It appears that its effect on the disc is negligible compared to the effect of pressure, but it provides the dominant source of precession on the planet, competing with the precession induced by the gravitational potential of the disc. For instance, in the 3D adiabatic case, the precession rate of the planet-dominated mode is $4.4\times 10^{-7}$ without the GR term, while it is $1.0\times 10^{-6}$ when GR is included.

\begin{table}
  \begin{tabular}{lcrr}   
  \multicolumn{1}{c}{Model} & \multicolumn{1}{c}{Mode order} & \multicolumn{1}{c}{precession rate} & \multicolumn{1}{c}{growth rate}  \\ 
\hline 
3D adiabatic &      0 & $ 8.369\times 10^{-4} $ & $ 3.470\times 10^{-4} $ \\
 &   1 & $ 3.156\times 10^{-5} $ & $ 1.035\times 10^{-5} $ \\
 &   3 & $ 1.553\times 10^{-6} $ & $ 4.533\times 10^{-7} $ \\
 &   P & $ 1.005\times 10^{-6} $ & $ 9.298\times 10^{-9} $  \\ 
  \hline
2D adiabatic &    0 & $-1.862\times 10^{-6} $ & $ 2.981\times 10^{-9} $ \\
 & 1 & $-1.086\times 10^{-5} $ & $ 6.920\times 10^{-8} $ \\
 & 7 & $-1.681\times 10^{-4} $ & $ 1.971\times 10^{-5} $ \\
 & P & $ 9.700\times 10^{-7} $ & $ 2.585\times 10^{-7} $ \\
\hline
3D isothermal &      0 & $ 1.064\times 10^{-3} $ & $ 2.985\times 10^{-4} $ \\
 & 1 & $ 3.944\times 10^{-5} $ & $ 1.067\times 10^{-5} $ \\
 & 2 & $ 1.240\times 10^{-6} $ & $ 3.999\times 10^{-7} $ \\
 & P & $ 9.524\times 10^{-7} $ & $ 1.280\times 10^{-8} $ \\
\hline
2D isothermal &	     0 & $-2.504\times 10^{-6} $ & $ 6.683\times 10^{-9} $ \\
 & 1 & $-1.533\times 10^{-5} $ & $ 7.468\times 10^{-5} $ \\
 & 2 & $-9.292\times 10^{-6} $ & $ 8.871\times 10^{-8} $ \\
 & P & $ 7.920\times 10^{-7} $ & $ 1.388\times 10^{-7} $ \\
\hline
  \end{tabular}
  \caption{Precession rates and growth rates for the lowest-order modes of our fiducial example, when viscosity and secular and resonant interactions with the planet, and back-reaction on the planet are included, in the 2D and 3D adiabatic and isothermal cases. The mode that is dominated by the planet is labelled by ``P''. The shape of the corresponding modes is shown in Fig.~\ref{fig:fiduPPVRPGR}.}
  \label{tab:fiduPPVRPGR}
\end{table}

\subsection{Self-gravity}
\label{sec:fidusg}

\begin{figure*}
    \begin{center}
    \includegraphics[width=2\columnwidth]{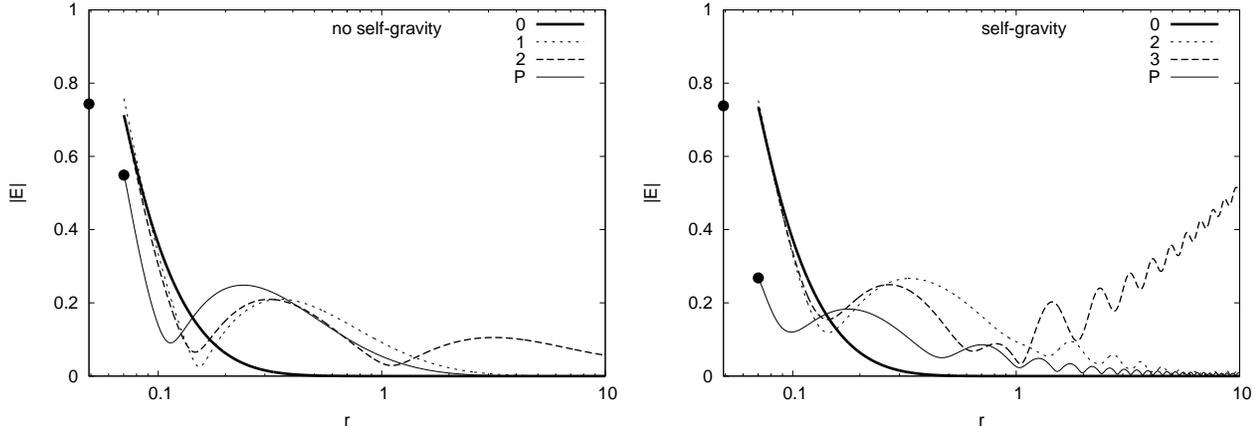}
    \caption{Shape of a few representative modes for a viscous disc, including secular and resonant interactions with a planet, and the back-reaction on the planet, in the 3D adiabatic case, without (left panel) and with (right panel) self-gravity. The number in the top right corner indicates the order of the corresponding mode. The mode labelled ``P'' is the planet-dominated mode, and the eccentricity of the planet with respect to that of the disc is marked by a black dot. Their precession rates are given in Table \ref{tab:fiduPPVRPSG}.}
    \label{fig:fiduPPVRPSG}
    \end{center}
\end{figure*}

Finally, we add self-gravity to the disc. In our fiducial example with $\qd=0.005$, the effect of self-gravity is negligible, so we arbitrarily multiply it by a factor of 100 in this section. In Fig.~\ref{fig:fiduPPVRPSG} and Table \ref{tab:fiduPPVRPSG} we show, for the 3D adiabatic model only, how self-gravity affects the modes when it is included compared to when it is not. Without self-gravity, the disc modes are not significantly affected when the mass of the disc is increased, but the growth rate of the planet-dominated mode is significantly increased, since it depends on the mass of the disc. The main effect of adding self-gravity is to cause more modes to precess progradely in the disc. There still exists a zero-order confined mode that is equivalent to the one that is found when self-gravity is not included. However the higher-order disc modes are affected by self-gravity. Most notably, there is now a low-order mode that propagates outwards in the disc, with a large AMD in its outer part, and hence a low growth rate. 

The planet-dominated mode remains mostly unchanged, although its growth rate is slightly lower with the inclusion of self-gravity. While the planet mode decays to zero in the inner part  of the disc without self-gravity, it now becomes oscillatory, and therefore carries more AMD, reducing its growth rate. 

In this example we have chosen a high value of surface density, perhaps unrealistic, in order to highlight the role of self-gravity. For more realistic surface density values, we find that self-gravity does not play an important role in the eccentricity evolution.

\begin{table*}
  \begin{tabular}{crrcccrr}   
     & \multicolumn{1}{c}{No self-gravity} & & & & & \multicolumn{1}{c}{Self-gravity} &  \\ 
     \hline
   \multicolumn{1}{c}{mode order} & \multicolumn{1}{c}{precession rate} & \multicolumn{1}{c}{growth rate} & & & \multicolumn{1}{c}{mode order} & \multicolumn{1}{c}{precession rate} & \multicolumn{1}{c}{growth rate} \\ 
\hline 
0 & 8.393$\times 10^{-4}$ &  3.551$\times 10^{-4}$ & & & 0  & 9.651$\times 10^{-4}$ & 3.548$\times 10^{-4}$  \\
1 & 2.952$\times 10^{-5}$ &  2.577$\times 10^{-6}$ & & & 2  & 1.852$\times 10^{-4}$ & 4.077$\times 10^{-6}$ \\ 
2 & 1.457$\times 10^{-6}$ &  2.717$\times 10^{-7}$ & & & 3  & 7.745$\times 10^{-5}$ & 1.450$\times 10^{-8}$  \\
P & 4.979$\times 10^{-5}$ &  1.521$\times 10^{-5}$ & & & P  & 5.103$\times 10^{-5}$ & 1.005$\times 10^{-5}$  \\
\hline
  \end{tabular}
  \caption{Precession rates and growth rates for the lowest-order modes of our fiducial example for a viscous, self-gravitating disc, including secular and resonant interactions with a planet, and the back-reaction on the planet, in the 2D and 3D abiabatic and isothermal cases. The mode that is dominated by the planet is labelled by ``P''. The shape of the corresponding modes is shown in Fig.~\ref{fig:fiduPPVRPSG}.}
  \label{tab:fiduPPVRPSG}
\end{table*}

\section{Influence of various parameters}
\label{sec:param}

Using a complete description of the disc--planet interactions that encompasses all the effects described above, we explore the behaviour of the precession rate and growth rate when varying different physical parameters.

In what follows, we study only the adiabatic case, in 2D and 3D. We identify the mode in the disc with the largest growth rate, which we label ``disc mode'' in all figures. We also characterize the evolution of the planet, by identifying the growing mode whose AMD is mostly located in the planet. We label this mode the ``planet mode''.

Our parameter survey is based on the previous fiducial values:
$\Sigma_0=1.3\times 10^{-4}$ (i.e. $\qd=0.005$), $\rin=0.07$,
$\rout=10$, $h_0=0.05$, $\ab=0.004$, $\ap=0.049$ (such that
$\ap/\rin=0.7$) and $\qp=0.003$.  In each case, we allow one parameter
to vary and discuss how it affects the precession and growth rates.

\subsection{Mass of the planet}
\label{sec:paramqp}

\begin{figure*}
    \begin{center}
    \includegraphics[width=2\columnwidth]{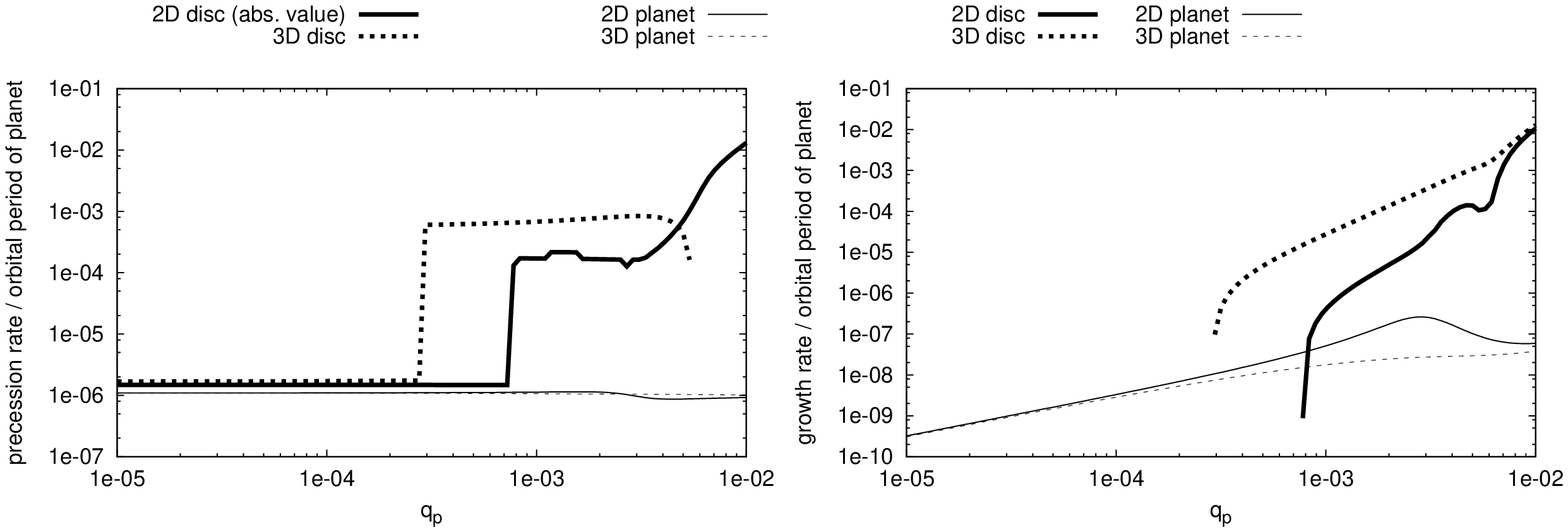}
    \caption{Precession rate (left) and growth rate (right) for the disc and planet modes with the highest growth rate, as functions of the planet-to-star mass ratio $\qp$. In 2D, the disc mode has a retrograde precession and we plot its absolute value. In 3D, the disc mode is mostly prograde, but becomes retrograde when $\qp\gtrsim 5\times10^{-3}$, and therefore does not appear on this logarithmic scale plot. Similarly, the disc mode is decaying for $\qp\lesssim 3\times 10^{-4}$ (resp. $8\times 10^{-4}$) in 3D (resp. 2D), and is not shown here.}
    \label{fig:paramqp}
    \end{center}
\end{figure*}

By analogy with the Schr\"odinger equation, we expect that the more massive the planet is, the deeper the effective potential well near the inner edge of the disc will be. This, in turn, can help the confinement of a mode in the inner parts of the disc, with small AMD, and hence a large growth rate. In this Section we explore the variation of the precession and growth rate as functions of the planet mass. Results are shown in Fig.~\ref{fig:paramqp}.

In the disc, 3D effects allow for the confinement of a mode, but the growth of this mode is determined by the balance between resonant effects and viscosity. For low-mass planets, viscosity prevents the resonant growth. This explains why, in Fig.~\ref{fig:paramqp}, the growth of the eccentric mode in the 3D disc becomes possible only when $\qp \gtrsim 3\times 10^{-4}$. In 2D, the growth only happens for $\qp \gtrsim 8\times 10^{-4}$ with our choice of parameters. When the mass of the planet is too low, the main contribution to the growth rate (Eq. [\ref{eq:intstot}]) comes from the viscous term (Eq. [\ref{eq:jvisc}]) which prevents damping. In addition, a small planet mass prevents the confinement of a mode in 2D. Only when the planet becomes sufficiently massive can the mode become significantly trapped, which in turn favours the growth. For very massive planets, the 2D mode is trapped in the deep potential created by the planet, and can grow on a timescale similar to the 3D case.

As we have already noted in Section \ref{sec:fiduplanet}, the planet-dominated mode experiences a faster growth in 2D than in 3D, but its growth is overall much smaller than the growth of disc-dominated modes.

In 2D, the fastest growing disc-mode is retrograde, and we plot the absolute value of its precession rate in Fig.~\ref{fig:paramqp}. The 3D disc supports a series of low-order prograde modes that are confined in the inner parts of the disc. An additional interesting feature appears in the 3D case: when $\qp\gtrsim 0.005$, the fastest growing mode becomes retrograde (and therefore is not plotted in Fig. \ref{fig:paramqp}, which is in logarithmic scale). Such modes, which we already discussed in 2D in Section \ref{sec:fiduvr}, are in apparent contradiction with the results of Section \ref{sec:schrod}, which predicted that the confined mode should be prograde. Confined retrograde modes are purely a consequence of the resonant growth of eccentricity. When the planet's mass is large enough, the growth rate becomes faster than the precession rate. Physically, this corresponds to a mode that grows faster in the vicinity of the planet than it propagates towards the outer parts of the disc. Such 3D retrograde modes however are rather rare, and in most cases the 3D confined growing mode is progade.

\subsection{Mass of the disc}
\label{sec:paramqd}

One can expect that, when self-gravity is negligible, the precession and growth rates of the eccentricity of the disc do not depend on the disc mass. Indeed, the evolutionary equations (\ref{eq:2dadia})--(\ref{eq:3diso}), or equivalently the effective potentials given in equations (\ref{eq:schrodiso}) and (\ref{eq:schrodadia}), are independent of the scaling factor $\Sigma_0$ that determines the total mass of the disc, provided that the contribution of the planet to the mode is negligible. 

For the planet mode, since most of the AMD is contained in the planet, we have $A\sim |\Ep|^2\Mp\ap^2\Op/2$, which is independent of $\Sigma_0$, while all contributions to the growth rate are proportional to $\Sigma_0$, so we expect the growth rate to scale linearly with the disc mass (see Eq. \ref{eq:intstot}). 
Regarding the precession rate of the planet-dominated mode, the secular contribution from the disc also scales linearly with disc mass (see Eq. \ref{eq:intpd}). At low disc mass, the precession rate of the planet is only weakly affected by the secular contribution from the disc. The main contribution in this regime comes from general relativity (and other short-range forces if they were included), which is independent of disc mass (see Eq. \ref{eq:intsrf}).

\begin{figure*}
    \begin{center}
    \includegraphics[width=2\columnwidth]{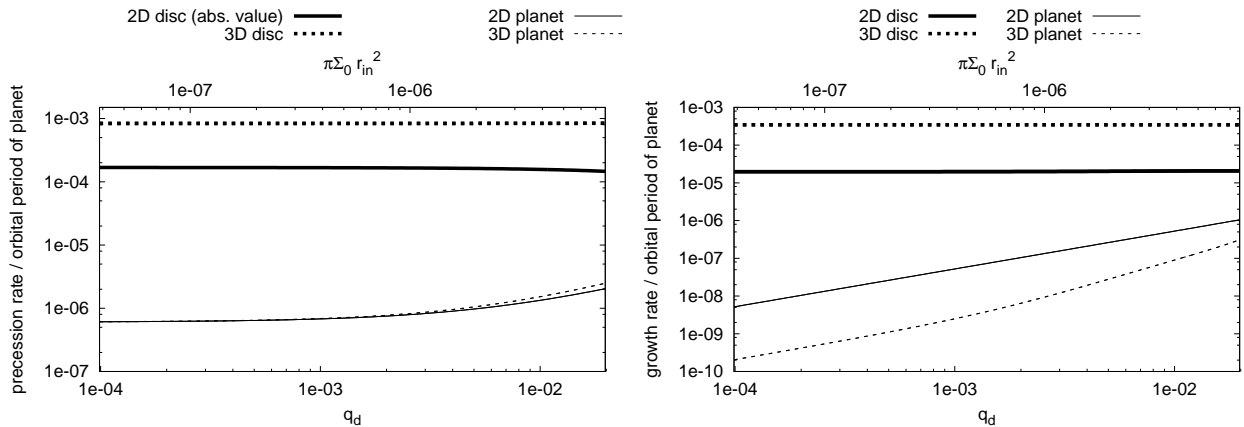}
    \caption{Precession rate (left) and growth rate (right) for the disc and planet modes with the highest growth rate, as functions of the disc-to-star mass ratio $\qd$. In 2D, the disc mode has a retrograde precession and we plot its absolute value. On the top $x$-axis we show the local disc mass near the planet $\pi\ssc\rin^2$ that corresponds to the total mass of the disc.}
    \label{fig:paramqd}
    \end{center}
\end{figure*}

The behaviours described above can be seen in Fig.~\ref{fig:paramqd}, where the precession and growth rates are plotted against the disc-to-star mass ratio $\qd$. There exists a 3D disc mode that precesses progradely and independently of disc mass (an equivalent 2D mode exists, but has a retrograde precession rate which does not appear on this logarithmic plot). As explained above, the precession rate of the planet mode is a combination of general relativity (independent of the disc mass) at low disc mass and secular effects from the disc (proportional to the disc mass) at high disc mass.
There exists a growing eccentric mode in the disc, which grows much more rapidly in 3D than in 2D. This is mostly due to our choice of parameters: the planet is not massive enough and close enough to the inner edge of the disc to allow for the confinement of a mode in the inner part of the disc in 2D. Both in 2D and 3D, the disc modes show no dependence on disc mass, as expected. The planet modes scale linearly with the disc mass, as was also expected. As discussed at the end of Section \ref{sec:paramqp}, there is a tendency for the 2D planet mode to contain less AMD than its 3D counterpart, and therefore grow more rapidly. 

The total mass of the disc is not necessarily an appropriate measurement of the perturbation induced by the disc on the planet. It is the local mass of the disc near the inner edge that mostly interacts with the planet. It can be defined as $\pi\Sigma_0\rin^2$, the amount of mass near the inner edge of the disc if its surface density was not decaying to zero. In Fig.~\ref{fig:paramqd} we also show this quantity for a given disc mass on the top $x$-axis. For the disc masses considered here, the local inner mass of the disc is lower than the mass of the planet. In addition, self-gravity is not expected to play an important role for these discs. In Section \ref{sec:rice}, where we compare our results with existing simulations in the literature, we will see an example where the local mass in the inner disc is larger than the mass of the planet, and how this leads to a mode switching.

\subsection{Location of the planet}
\label{sec:paramap}

The theory of disc--planet interactions suggests that, once a planet enters the central cavity of the disc, it will stop migrating when the torque exerted by the Lindblad resonances vanishes, as the resonances all fall into the clean gap. In this Section we investigate the influence of the orbital radius of the planet on the precession and growth rates of eccentric modes. 

The semi-major axis of the planet is expected to be of primary importance regarding the possibility for eccentric modes to grow in a disc. In addition to determining how many resonances are located in the disc, it plays a major role in shaping the potential well in which a mode can be confined, especially in 2D (see Eq. \ref{eq:schrodiso}--\ref{eq:schrodadia}).

\begin{figure*}
    \begin{center}
    \includegraphics[width=2\columnwidth]{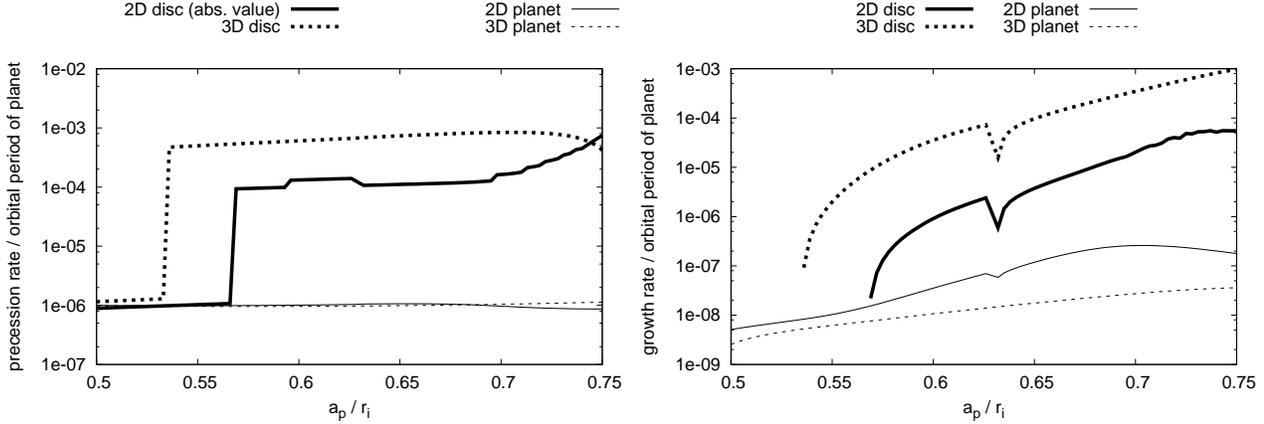}
    \caption{Precession rate (left) and growth rate (right) for the disc and planet modes with the highest growth rate, as functions of $\ap/\rin$. In 2D, the disc mode has a retrograde precession and we plot its absolute value. Similarly, the disc mode is decaying for $\ap/\rin\lesssim 0.53$ (resp. $0.57$) in 3D (resp. 2D), and is not shown here.}
    \label{fig:abgrt}
    \end{center}
\end{figure*}

The precession and growth rates of the fastest growing mode are displayed in Fig.~\ref{fig:abgrt}, as functions of $\ap/\rin$. We choose this variable as it help visualizing the effect of resonances. We recall that we have chosen $\rin=0.07$ au, and our fiducial example described above has $\ap/\rin=0.7$. In 3D, when the planet is far enough from the disc, eccentricity evolution is dominated by viscous damping, which prevents any mode to grow. Only when the planet starts getting closer to the disc can the 1:3 ELR eventually overcome viscous damping and give rise to eccentricity growth. However, as discussed in Section \ref{sec:fiduvr}, the amplitude of the contribution from the 1:3 ELR is rather weak, and the overall growth rate is small. As the planet gets closer to the disc's inner edge, the 2:4 ELR and 3:5 ELR become the major source of eccentricity growth, while the 1:2 ECR damps it. The dip in eccentricity growth seen at $\ap/\rin \sim 0.64$ corresponds to the location where the 1:2 ECR is most effective at damping AMD (we prefer to talk about AMD rather than eccentricity here, since it is the eccentricity weighted by the surface density profile that really matters for the growth of the mode). 

It might seem puzzling at first that such a growing mode exists and that ECRs are rather ineffective at damping it. We find that the resonant widths of both ELRs and ECRs play a major role in this process. As we have noted in Section \ref{sec:res}, ELRs are better modelled if their centre is shifted by one resonant width in the direction away from the planet. Hence, ELRs for which the nominal radius is in the cavity can actually operate quite strongly in the disc. From Table \ref{tab:rab}, one would expect that, if $\ap/\rin=0.7$, only the 1:3 ELR would be located deeply in the disc, with the 2:4 ELR just at the edge of the disc, and all the other resonances in the cavity. However, allowing all the ELRs to be shifted and to have a finite width enables for the 2:4, 3:5, 4:6 and possibly higher order ELRs to sit in the disc. On the other hand, ECRs do not have this shift, and have a very narrow width. Hence, only the 1:2 resonance can operate in the disc, and it does not provide enough damping to strongly affect the net growth rate. According to Fig.~\ref{fig:abgrt}, only when the planet orbits at $\ap/\rin\approx 0.64$ can the growth rate be significantly reduced. This is not surprising as it corresponds to the radius at which the 1:2 ECR starts being located inside the disc. As the planet gets closer to the disc, more and more ELRs enter the disc and progressively make the 1:2 ECR ineffective, recovering the result where only ELRs are taken into account. Hence, even without saturation of the corotation torque, eccentric modes can grow because of eccentric Lindblad resonances.

In 2D, the growing mode appears when the planet is slightly closer to the inner edge of the disc, compared to the 3D case. As we have seen in Section \ref{sec:schrod}, it is much harder to trap a mode in a deep potential well in 2D than in 3D. In fact, while a 3D disc naturally supports a set of confined prograde modes, the equivalent mode in 2D can only be found when a planet is close enough to the disc (and/or massive enough) to allow a relative confinement of the mode near the inner edge, which in turns allows the mode to grow. We have also run a similar calculation with $\qp=0.001$, all other parameters being equal, and we found that a growing mode in 2D only appeared for $\ap/\rin\gtrsim 0.67$. Reducing the planet's mass decreases the depth of the effective potential well, so that only a planet close to the inner disc can help trap a mode in 2D.

The precession rate of the planet mode is always positive (prograde mode) and only weakly depends on the planet's semi-major axis. The disc mode has a negative precession rate in 2D (retrograde mode) and we plot its absolute value in Fig.~\ref{fig:abgrt}. It precesses progadely in 3D, with a significant jump in precession rate as soon as the mode is allow to grow.

These results depend on the value we adopt for the resonant width. We explore this dependence in Section \ref{sec:paramw}.

\subsection{Disc aspect ratio}
\label{sec:paramh}

\begin{figure*}
    \begin{center}
    \includegraphics[width=2\columnwidth]{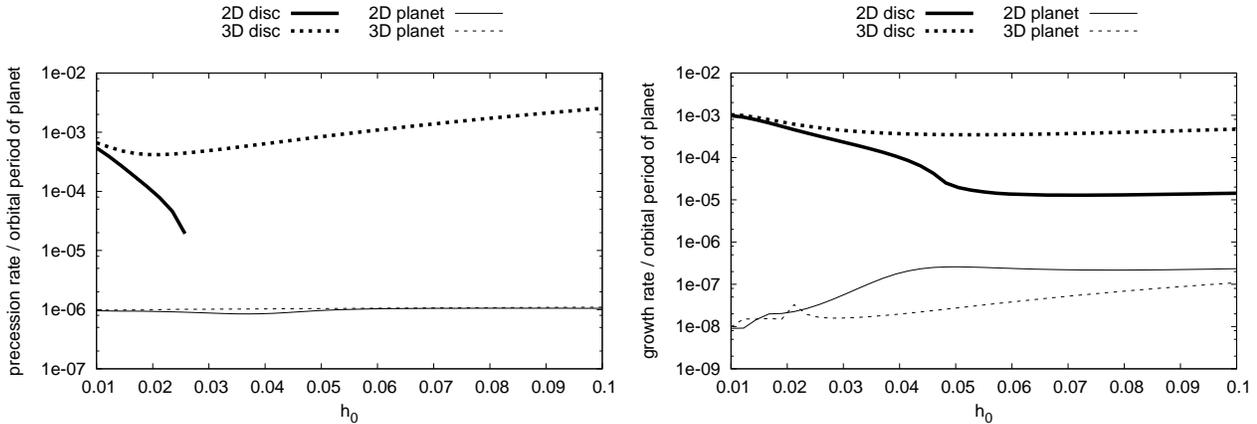}
    \caption{Precession rate (left) and growth rate (right) for the disc and planet modes with the highest growth rate, as functions of $h_0$. In 2D, the disc-dominated mode becomes retrograde for $h_0\gtrsim0.025$ and does not appear on the logarithmic plot.	}
    \label{fig:hgrt}
    \end{center}
\end{figure*}

We vary the disc aspect ratio $h_0$ in Fig.~\ref{fig:hgrt}. How the precession and growth rates should vary with $h_0$ is not obvious from Section \ref{sec:int}.  Several contributions to the precession rate are proportional to pressure and therefore to $h_0^2$; however, these integrals also depend on the mode shape, which in turn depends in a non-trivial way on $h_0$.  The aspect ratio also appears indirectly in the ELR width (see equation \ref{eq:rwelr}). 

The precession rate of a 2D disc shows that for small $h_0$ the fastest growing mode in the disc has a progade precession, which decreases as $h_0$ increases, until it becomes negative. In 2D the pressure-induced precession scales as $h_0^2$, such as for thin discs, the precession rate is dominated by the prograde contribution coming from the planet. This prograde contribution is eventually overcome by the pressure term when the disc becomes sufficiently thick, and the precession is retrograde. When $h_0$ is small, the potential well in which the mode is trapped arises mainly from the planet, with little contribution from pressure. In this regime, the growth rates in 2D and 3D are therefore similar.
 
At large $h_0$, the growth rate in 2D is significantly smaller than in 3D. A similar calculation but with $\qp=0.001$ instead of 0.003 showed that the growth of a 2D mode was completely suppressed for $h_0\gtrsim0.065$. The 2D adiabatic case does not allow generally for a bound state to be supported in the inner disc. Only the presence of a planet can, in some cases, deepen the potential well and allow for a trapped mode. This can be understood from the effective potential that we derived in Eq. (\ref{eq:schrodadia}). In this equation, the last term in the effective potential $V(x)$ shows the competition between the potential well of the planet and the disc aspect ratio (where we recall that $\epsilon=h_0$). When $h_0$ becomes large enough, it causes the term due to the planet to become ineffective, and no bound state can be supported. 

In the 3D case, the growth rate of the disc-dominated mode first decreases before increasing again. As in the 2D case, the decrease can be interpreted as a consequence of the weaker confinement of the mode by the planet as $h_0$ increases. However, the width of the ELRs is an increasing function of $h_0$, which in turns allows for a larger growth rate, as a larger part of the disc contributes to the growth. This explains why the growth rate slightly increases again at large $h_0$.

\subsection{Viscosity}
\label{sec:paramvisc}

From the integral relations derived in Section \ref{sec:int} (see eq. \ref{eq:jvisc}), one can expect the viscous damping of eccentricity to scale linearly with $\ab$. We have set up a calculation in which $\ab$ varies from $10^{-4}$ to $10^{-1}$, not shown here. We find that the scaling is indeed linear. 
It is important to remember that we are using an effective bulk viscosity in this paper, representing whichever thermal or mechanical processes (apart from resonant effects) damp the eccentricity. While it provides a convenient formulation of eccentricity damping in a linear theory, it does not necessarily represent the effective shear viscosity that is driving accretion.
In any case our results indicate that the bulk viscosity is not strong enough to oppose the resonant growth of eccentricity. Only in the case where the 1:3 ELR is the only resonance in the disc can viscosity compete with it (and possibly suppress the growth rate).

In the case where the planet was orbiting in a gap instead of a cavity, viscosity would presumably play an important role in setting the gap's shape, and therefore in the surface density profile. It is not the case here, since we consider a planet in a cavity, whose origin is supposed to be the truncation of the disc by the stellar magnetosphere. 

\subsection{Resonant width}
\label{sec:paramw}

In Section \ref{sec:res} we have allowed the resonances to spread over a few rings by giving them a finite width. As discussed in Section \ref{sec:paramap}, it is the resonant width of ELRs that allows for a large number of ELRs to sit in the disc and contribute positively to the growth rate. However, there are some uncertainties on the exact value of the width, for both ELRs and ECRs. Estimating the influence of the resonant width is made difficult by the parametrization we used in section \ref{sec:res} (see equations \ref{eq:rwelr} and \ref{eq:rwecr}): each resonant width depends on the wavenumber $m$, and the ELRs' width decreases with increasing $m$, while it increases for ECRs. 

\begin{figure*}
    \begin{center}
    \includegraphics[width=2\columnwidth]{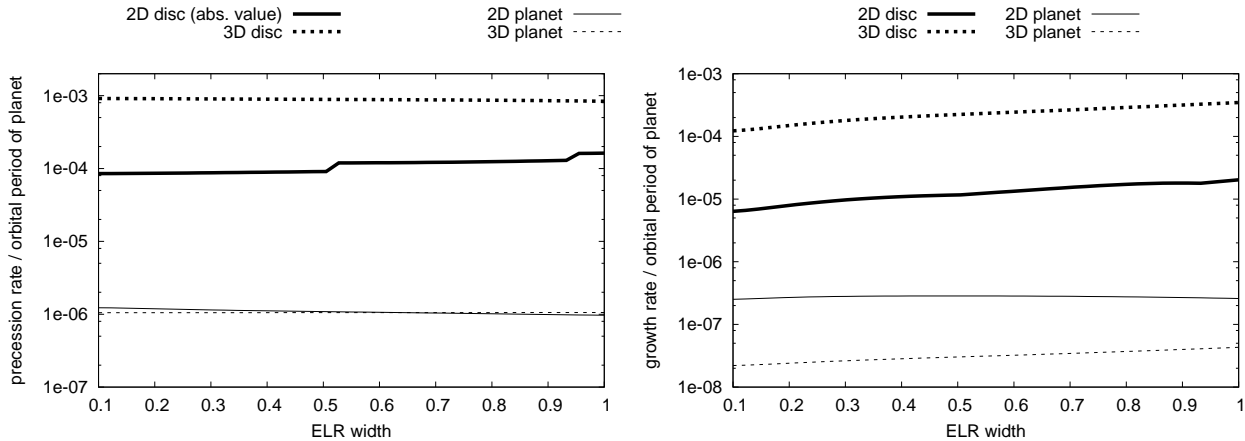}
    \caption{Precession rate (left) and growth rate (right) for the mode with the highest growth rate, as functions of the ELR resonant width. 
    }
    \label{fig:rw1em6}
    \end{center}
\end{figure*}

In order to estimate how the growth rates might be affected by the resonant width, we adopt the following parametrization for ELRs:
\begin{equation}
\label{eq:rwelr2}
\frac{w_{\rm L}}{r}\bigg|_{\rm ELR} = \lambda \left( \frac{h_0^2}{3(m+1)} \right)^{1/3},
\end{equation}
and we vary $\lambda$. In Fig.~\ref{fig:rw1em6}, we show the growth rates for $0.1<\lambda<1$. For ECRs, we have varied the value of $e$ in eq. (\ref{eq:rwecr}) from $10^{-6}$ to $10^{-2}$ and checked that it has no significant effect on the growth rate, as it is always so narrow that only the 1:2 ECR lies in the disc. We do not show the effect of varying the ECR width here, as the plots would be similar to Fig.~\ref{fig:rw1em6}. As the width of ELRs increases, a larger part of the disc contributes to the growth of eccentricity, while the precession rates remain unaffected. 

Although Table \ref{tab:rab} suggests that resonances with higher $j$ should be stronger, this is not necessarily the case. Indeed, these resonances lie close to the planet, and according to eq. (\ref{eq:rwelr}), their width decreases as $j$ increase, and they are not broad enough to affect the disc. In addition, we recall that we have applied a cutoff so that resonances with $j\gtrsim r/H$ are not included.

Note that in Section \ref{sec:res}, we have derived the resonant width for ELRs from a dispersion relation. In principle, this dispersion relation is modified when self-gravity is included. We find that the resonant width is now approximately given by the solution of the equation
\begin{equation}
3(m+1)\frac{w_{\rm L}^3}{r^3} + 2\pi\frac{\Sigma(\rres)\rres^2}{M_*}\frac{w_{\rm L}}{r} = h_{0}^2.
\end{equation}
The additional term due to self-gravity will be zero when $\Sigma(\rres)=0$ (i.e., when the resonant radius is not in the disc). In addition, when local disc mass at the resonant location $\pi\Sigma(\rres)\rres^2$ is small, the addition of this new term brings a negligible modification to the resonant width, and can be ignored. Note that we have neglected the subtle effect caused by the shifting of the resonance location away from the nominal resonant radius $\rres$. This effect does not change the general conclusion that the effect of self-gravity on the resonant width is negligible.

\section{Discussion}
\label{sec:discu}

\subsection{Comparison with Rice et al. 2008}
\label{sec:rice}

\citet{rice08} suggested that the paucity of hot Jupiters could be caused by their interactions with the inner disc as they orbit in the stellar magnetospheric cavity. They found that, for a wide range of parameters, eccentricity can grow and eventually lead to collision with the star. In order to speed up their calculations, they increased the mass of the disc, arguing that it would simply change the timescale on which the eccentricity grows, and that everything can be rescaled linearly. They choose a surface density of $10^{-2}$ in their units, which corresponds to $10^{4}$ times the expected value for such a disc.

We set up a calculation similar to theirs, of a 2D locally isothermal disc with $\Sigma(r)\propto r^{-1}$ (but we keep the two taper functions which cause $\Sigma$ to be zero at each edge). We take $h_0=0.05$, $\ab=10^{-3}$, $\rin=1$, $\rout=10$, $\ap=0.7$ and $\qp=0.005$. We let the mass of the disc vary to cover the range from both physical values to the unrealistic (yet convenient) values of \citet{rice08}. The calculation does not include self-gravity or relativistic precession.

\begin{figure}
    \begin{center}
    \includegraphics[width=1\columnwidth]{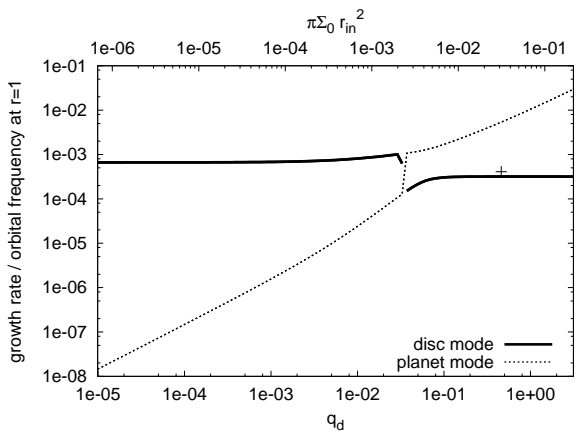}
    \caption{Growth rate of the fastest growing disc mode (solid line) and planet mode (dotted line), in a calculation similar to that of Rice et al. (2008). The bottom $x$-axis represents the disc-to-star mass ratio $\qd$, and the equivalent local disc mass near the planet $\pi\Sigma_0\rin^2$ is represented on the top axis. The cross shows the growth rate of the planet's eccentricity as found by \citet{rice08}.}
    \label{fig:rice}
    \end{center}
\end{figure}

In Fig.~\ref{fig:rice} we plot the growth rates of the disc-dominated and planet-dominated modes, as a function of the mass of the disc $\qd$, and as function of the more meaningful quantity, the local disc mass near the planet $\pi\Sigma_0\rin^2$. Since in this calculation $\rin=1$, it also gives a direct measure of $\Sigma_0$. Note that for a better comparison with the results of \cite{rice08}, we give here the growth rate divided by the orbital period at $r=\rin=1$. 

For $\Sigma_0=10^{-2}$, the growth rate is about $4\times 10^{-3}$ orbits at $r=1$, a value 10 times larger than what was found by \citet{rice08}. This discrepancy could arise from the different boundary conditions at the inner radius in the two calculations (an outflow in \citet{rice08} and a free edge in our calculation). It is also possible that a 5 Jupiter-mass planet at $\ap=0.7$ experiences inward migration. We have repeated the calculation with a 1 Jupiter mass planet, which shows no migration in \citet{rice08}, but still find a higher growth rate than the authors (although the growth rate for a 1 Jupiter mass planet is only poorly constrained in their work). 

The scaling of growth rate with disc mass is linear for a wide range of masses, as claimed by the authors. However, for a disc with real physically motivated parameters, $\Sigma_0$ would be in the range of $\sim 10^{-6}$ in the units used for this calculation. At this point the growth rate of the planet is lower than the growth rate of eccentricity in the disc by 5 order of magnitudes. Therefore, in a real disc--planet system of this kind, one expects the eccentricity to grow much more rapidly in the disc-dominated mode than in the planet-dominated mode. 
Here our linear model fails to predict what would happen to the coupled system in the limit of large eccentricities. Presumably the disc mode will eventually undergo a non-linear saturation. One could speculate that, as the disc possesses a large reservoir of AMD, it could slowly and chaotically diffuse it in the planet, therefore exciting its eccentricity in a process similar to secular chaos in planetary systems \citep{wl11}. If the chaotic behaviour finds its origin in the overlapping of two secular resonances \citep[see, e.g.,][]{lw11}, it may be necessary to invoke the presence of two different eccentric modes in the disc, or additional planets. However it is possible that the timescale of such non-linear interaction might be longer than the disc lifetime.

Finally, we remark that the fastest growing mode in the system switches from being a disc-dominated mode to a planet-dominated mode when the local mass in the disc is of the order of the mass of the planet. In the regime where the two masses are nearly equal, there exists a strong coupling between the planet and the disc, where they share an almost equal amount of AMD. Away from this mode transition region, the scaling of mode with growth rate is roughly linear.

\subsection{Disc model}

Our disc model described in Section \ref{sec:discmodel} is based on a strong assumption regarding the opacity in the inner parts of the disc. We have conducted a series of runs in which the disc is described using an $\alpha$-disc model for which the surface density and aspect ratio read
\begin{equation}
\Sigma = \Sigma_{0}\left(\frac{r}{r_{0}}\right)^{-1/2} \left(1 - \sqrt{\frac{\rin}{r}}\right)^{7/10}\tanh\left(\frac{\rout-r}{\wout}\right),
\end{equation}
and
\begin{equation}
\label{eq:hoverrb}
\frac{H}{r} = h_{0}\left(1-\sqrt{\frac{\rin}{r}}\right)^{3/20} \left( \frac{r}{\rin}\right)^{1/8}.
\end{equation} 
Such a model implies a slight flaring of the disc with
radius. We found that the overall results described in this paper
remain unchanged with this description. A mode can still be trapped in
the inner part of the disc, which grows on the same timescale, and
shares the same properties regarding the difference between 2D and 3D
models.

We have chosen $\gamma = 7/5$ in the adiabatic case, but checked that results remain similar for other values (e.g., $\gamma=5/3$). Non-adiabatic effects could in principle affect the eccentricity growth or decay, but are beyond the scope of this paper.

Finally, the lowest-order mode is fairly independent of the outer radius $\rout$, since it decays rapidly with radius. Higher-order trapped prograde modes are expected to decay to zero at large radii, provided the disc's outer radius is large enough, and should be independent of $\rout$ as $\rout\to\infty$. However they usually contain more AMD and thus grow less rapidly than the lowest-order confined mode. Global retrograde modes will depend on our choice of $\rout$.

\section{Conclusion}
\label{sec:conclu}

In this paper we have formulated a system of linear equations governing the evolution of small eccentricities in systems consisting of one or more planet coupled to a gaseous disc through secular and resonant interactions. We have applied this formalism to the problem of a planet orbiting in the cavity of a protoplanetary disc, under the influence of various physical effects. 

The inclusion of 3D effects allows for the existence of an eccentric mode trapped in the inner parts of the disc. The existence of this mode is very robust over a large range of parameters. The distribution of AMD in this mode allows for a rapid growth rate, compared with 2D models. For a wide range of disc and planet parameters, an eccentric mode can grow in the disc on a timescale smaller than the disc's lifetime, although usually beyond the scope of current direct numerical simulations of disc--planet systems.

We find that allowing for a finite resonant width and shifting the resonance away from its nominal radius leads to several ELRs operating in the disc. It is mostly the 2:4 and 3:5 ELRs that dominate the growth rate, with a small contribution from the 4:6. The 1:3 ELR is found to have little impact on the growth rate, because of its small amplitude. The only ECR that significantly contributes to the damping of eccentricity is the 1:2. Other ECRs might be present in the disc if the planet orbits closer to the inner edge, but we find that eccentricity readily grows nonetheless, because of the presence of several ELRs. This suggests that, even when the corotation resonances are not saturated, eccentricity can grow even for Jupiter-mass planets. If such growth was not observed in some numerical simulations, it is most likely because it occurs on (at least) thousands of orbital periods of the planet. In addition, growth in the disc occurs more rapidly than in the planet, and it is not clear whether the disc's eccentricity will undergo a nonlinear saturation, and how would it affect the planet.

We have compared our results with those of \citet{rice08}. For the same unrealistically massive disc, we also find that an eccentric mode can grow, which is dominated by the planet. \citet{rice08} argued that the growth rates could be scaled down linearly with the disc mass, and that massive planets could therefore experience eccentricity growth within the disc lifetime. However we caution that when scaling down the system, one enters the regime where the growth of eccentricity is dominated by a disc mode whose growth is much more rapid than the planet-dominated mode (Fig.~\ref{fig:rice}). It is not clear what the outcome of disc--planet interactions could be for a system with an eccentric disc. 

In this paper we have not considered planets orbiting in a gap, such as that created by massive planets undergoing type-II migration. This is because the gap profile is not easy to model in such case, and should evolve in time as the planet migrates. In addition, the planet is supposed to undergo type-II migration on a timescale that can be comparable to the growth rate of eccentricity. The assumption of a constant semi-major axis that is made here may no longer be a good approximation, and the contribution from the inner and outer discs will change as the planet comes closer to the star.

It remains to be understood how, as the disc--planet system evolves in time, AMD is redistributed between the two components. In the case where the disc contains most of the AMD, it is possible that, as the it becomes more and more depleted, AMD is given to the planet, allowing its eccentricity to grow. We remark that the same mechanism could potentially allow for inclination growth as well. This mechanism has important consequences, as it dictates the orbital architecture of the planetary systems at the beginning of the gas-free stage of planetary evolution.

\vskip 0.2 truein
--------

We thank Pin-Gao Gu for useful discussions about this problem, and Simon Goodchild for his contributions many years ago. We acknowledge support from STFC through grant ST/L000636/1.

\appendix

\section{Derivation of the eccentricity equations}
\label{app:ecceq}

\subsection{Linear theory in a three-dimensional inviscid disc}

\label{app:lineareq}

The derivation of the eccentricity equation is similar to that for the 2D inviscid disc as described by \citet{go06}. In cylindrical polar coordinates $(r,\phi,z)$, if the fluid has velocity components $(u,v,w)$, density $\rho$, pressure $p$ and orbits in a gravitational potential $\Phi$, then the fluid equations are
\begin{align} 
&\pd{u}{t} + u\pd{u}{r} + \frac{v}{r}\pd{u}{\phi} + w\pd{u}{z} - \frac{v^2}{r} = -\pd{\Phi}{r} - \frac{1}{\rho}\pd{p}{r},  \label{eq:du}\\ 
&\pd{v}{t} + u\pd{v}{r} + \frac{v}{r}\pd{v}{\phi} + w\pd{v}{z} + \frac{uv}{r} = -\frac{1}{r}\pd{\Phi}{\phi} - \frac{1}{r\rho}\pd{p}{\phi}, \label{eq:dv} \\ 
&\pd{w}{t} + u\pd{w}{r} + \frac{v}{r}\pd{w}{\phi} + w\pd{w}{z} = -\pd{\Phi}{z} - \frac{1}{\rho}\pd{p}{z}, \label{eq:dw} \\ 
&\pd{\rho}{t} + u\pd{\rho}{r} + \frac{v}{r}\pd{\rho}{\phi} + w\pd{\rho}{z} = -\rho\left[ \frac{1}{r}\pd{(ru)}{r} + \frac{1}{r}\pd{v}{\phi} + \pd{w}{z} \right]. \label{eq:drho}
\end{align}
We assume in this derivation that the disc is locally isothermal and therefore has the equation of state
\begin{equation}
p=\cs\rho,
\end{equation}
where the locally isothermal sound speed $c_{\rm s}(r)$ is a function of the radial coordinate only. This equation can be justified physically if the temperature of the disc is determined by the radiation of the central star, and any excess is rapidly radiated away by the disc.

In the basic state, representing a steady, circular disc, the velocity is $(u,v,w)=(0,r\Omega(r,z),0)$, and the density is $\rho(r,z)$. Therefore equations (\ref{eq:du}) and (\ref{eq:dw}) become
\begin{align}
& -r\Omega^2 = -\pd{\Phi}{r} - \frac{1}{\rho}\pd{}{r}(\cs\rho), \label{eq:dufo}\\
& 0 = -\pd{\Phi}{z} - \frac{1}{\rho}\pd{}{z}(\cs\rho). \label{eq:dwfo}
\end{align}
To solve these equations, each quantity is expanded as an asymptotic series in a small parameter $\epsilon^2\ll 1$, as in \citet{ogilvie01}. Here $\epsilon$ represents a characteristic value of $H/r$, where $H(r)$ is the semi-thickness of the disc. Each quantity $X$ is then written as
\begin{equation}
X=\epsilon^s(X_0+X_2\epsilon^2+\cdots),
\end{equation}
where the parameter $s$ is chosen in each case to provide appropriate scaling for the various quantities.

In a system of units such as $r$ and $\Omega$ are $O(1)$, the sound speed is $O(\epsilon)$, so we write $c_{\rm s} = \epsilon c_{\rm s0}$. We use a new coordinate $\zeta$ to describe the vertical structure of the disc, where $z=\epsilon\zeta$ so that $\zeta=O(1)$ within the disc. The potential can be split into two components, $\Phi_*$ due to the central star and $\Phi_{\rm p,d}$ due to the planet and the disc. As $\Phi_*$ is much larger than $\Phi_{\rm p,d}$, the latter is taken formally to be of $O(\epsilon^2)$. Using equation (\ref{eq:dufo}) at $O(1)$ gives the usual expression for the Keplerian angular velocity $\Omega_0=(GM_*/r^3)^{1/2}$, and we can use this to write the series for $\Phi_*=-GM_*(r^2+z^2)^{-1/2}$,
\begin{equation}
\Phi_* = -r^2\Omega_0^2 + \frac{1}{2}\Omega_0^2\epsilon^2\zeta^2 + O(\epsilon^4).
\end{equation}
Thus the vertical structure equation (\ref{eq:dwfo}) becomes
\begin{equation}
\label{eq:vertstruc}
\pd{\rho_0}{\zeta} = -\frac{\rho_0\Omega_0^2\zeta}{c_{\rm s0}^2},
\end{equation}
and the vertical density distribution is a Gaussian with scale height $H=c_{\rm s}/\Omega_0$,
\begin{equation}
\rho_0(r,\zeta) = \rho_0(r,0)\exp\left(-\frac{\Omega_0^2\zeta^2}{2c_{\rm s0}^2} \right).
\end{equation}
The angular velocity correction at $O(\epsilon^2)$ is given by equation (\ref{eq:dufo}) at $O(\epsilon^2)$,
\begin{equation}
-2r\Omega_0\Omega_2 = \frac{3\Omega_0^2\zeta^2}{2r} - \pd{\Phi_{\rm p,d}}{r} - \frac{1}{\rho_0}\pd{}{r}(c_{\rm s0}^2\rho_0).
\end{equation}

We then introduce perturbed quantities in the form $\Re[X'(r,z,t)\me^{-\im \phi}]$ for the velocity, density and potential. The angular form of the perturbations is chosen so they replicate the effect of a small eccentricity in the disc. The linearized, perturbed fluid equations are
\begin{align} 
&\pd{u'}{t} - \im\Omega u' - 2\Omega v' = -\pd{\Phi'}{r} -\cs\pd{}{r}\left(\frac{\rho'}{\rho} \right),  \label{eq:ldu}\\ 
&\pd{v'}{t}  - \im\Omega v' + \frac{u'}{r}\pd{}{r}(r^2\Omega) + w' \pd{}{z}(r\Omega) = \frac{\im \Phi'}{r} + \frac{\im\cs\rho'}{r\rho}, \label{eq:ldv} \\ 
&\pd{w'}{t} -\im\Omega w' = -\pd{\Phi'}{z} - \cs\pd{}{z}\left(\frac{\rho'}{\rho}\right), \label{eq:ldw} \\ 
&\pd{\rho'}{t} - \im\Omega\rho' + u'\pd{\rho}{r} + w'\pd{\rho}{z} = -\rho\left[ \frac{1}{r}\pd{(ru')}{r} - \frac{\im v'}{r} + \pd{w'}{z} \right]. \label{eq:ldrho}
\end{align}

Since we are interested in the evolution of the disc on a timescale much longer than the orbital timescale, all time derivatives are assumed to be of $O(\epsilon^2)$, so that $\partial_t\mapsto \epsilon^2\partial_T$ with a slow time variable $T$. The perturbed quantities can be expanded as series in $\epsilon^2$ in the same way as the steady-state quantities, with $u'$, $v'$ and $\rho'$ being $O(1)$ while $w'$ is $O(\epsilon)$ (we note that the overall normalization of the perturbations in a linear theory is arbitrary). Considering equations (\ref{eq:ldu}) and (\ref{eq:ldv}) at $O(1)$ we have, as in \citet{go06}, a general solution allowing velocities that describe a small eccentric perturbation expressed using the complex eccentricity $E=e\me^{\im\varpi}$,
\begin{equation}
\label{eq:uvp}
u_0'=\im r \Omega_0 E(r,T), \qquad v_0'=\frac{1}{2}r\Omega_0E(r,T). 
\end{equation}
The complex eccentricity could also depend on the vertical coordinate, but we neglect this possibility here. Physically, this requires a shear viscosity or some kind of stress to link the ellipses at different heights together \citep{lo06}.

In order to find an equation describing the slow time-evolution of the eccentricity, we consider equations (\ref{eq:ldu}) and (\ref{eq:ldv}) at $O(\epsilon^2)$. This expression includes the vertical velocity perturbation $w_0'$, so first an expression for this needs to be found. Substituting for $u_0'$ and $v_0'$ from equation (\ref{eq:uvp}) into equation (\ref{eq:ldrho}) produces an expression for the density perturbation,
\begin{equation}
\label{eq:rhop}
\rho_0'=r\pd{}{r}(\rho_0E) - \frac{\im}{\Omega_0}\pd{}{\zeta}(\rho_o w_0').
\end{equation}
Substituting this into equation (\ref{eq:ldw}) at $O(\epsilon)$, using equation (\ref{eq:vertstruc}) and simplifying produces a differential equation for $w'_0$,
\begin{equation}
\frac{\im c_{\rm s0}^2}{\Omega_0} \pd{^2 w_0'}{\zeta^2} - \im\Omega_0\zeta\pd{w_0'}{\zeta} = \Omega_0^2\left(3 + \dd{\ln \cs}{\ln r}\right)E\zeta.
\end{equation}
The solution is $w_0'=W(r,t)\zeta$ with
\begin{equation}
W = \im\Omega_0\left(3 + \dd{\ln \cs}{\ln r}\right)E.
\end{equation}

By taking a linear combination of equations (\ref{eq:ldu}) and (\ref{eq:ldv}) at $O(\epsilon^2)$, the second-order velocity perturbations can be eliminated and a differential equation found for $E$, using the relations above to eliminate $\Omega_2$, $\rho_0'$ and $w_0'$. We introduce the vertically integrated surface density $\Sigma_0$ defined as
\begin{equation}
\Sigma_0 = \int \rho_0 \id \zeta,
\end{equation}
where the integral is over the whole vertical extent of the disc. We then integrate the equation over the vertical extent of the disc, weighted by the density, This procedure can be justified if there is some viscosity in the disc, since the integration eliminates the viscous terms $\partial_\zeta(\rho_0\nu\partial_z u_2')$ and $\partial_\zeta(\rho_0\nu\partial_z v_2')$ that link the ellipses together as described above when we introduced the assumption that $E$ is independent of $\zeta$ \citep[cf.][]{ogilvie01}. Removing the scaling subscripts, we have the final form of the equation in the original variables,
\begin{equation}
\begin{split}
\label{eq:dedtapp}
\Sigma r^2 \Omega\pd{E}{t} & = \frac{\im}{2r}\pd{}{r}\left(\Sigma\cs r^3 \pd{E}{r} \right) + \frac{\im r}{2}\dd{}{r}(\Sigma\cs)E \\
& - \frac{\im}{2 r}\pd{}{r}\left(\Sigma \dd{\cs}{r}r^3 E \right) + \frac{3\im}{2r}\Sigma \dd{}{r}(\cs r^2)E \\
& - \frac{\im\Sigma}{2}\pd{}{r}\left(r^2\pd{\Phi_{\rm p,d}}{r} \right)E + \frac{\im\Sigma}{2r}\pd{}{r}(r^2\Phi_{\rm p,d}')E.
\end{split}
\end{equation}
Here $\Omega$ denotes the Keplerian angular velocity $\Omega_0$. If the disc is assumed to be two-dimensional then the equation is almost the same, but lacks the term $\frac{3\im}{2r}\Sigma \dd{}{r}(\cs r^2)E$, which arises from the three-dimensional structure and in particular from the three-dimensional term in the density perturbation equation (\ref{eq:rhop}).

\subsection{The perturbing potential}
\label{app:pertpot}

In a prototplanetary disc, the gravitational potential deviates from that of a point mass because of the presence of the planet(s) and the disc. We first derive the orbital-averaged perturbing potential arising from a planet. The result can then be extended to derive the potential arising from the disc, by treating the latter as a collection of annuli of matter. We evaluate the potential in the midplane $z=0$ since this is equivalent  to the lowest-order term in $\epsilon^2$ which is the only one to enter the eccentricity equation. 

The orbit-averaged potential of a planet can be calculated using the Gauss averaging method \citep[see, e.g.,][]{md99} in which the planet is smeared out along its orbit in such way that the mass along any part of the orbit is proportional to the time taken to traverse that part. If the planet is located at coordinates $(r_{\rm p}, \phi_{\rm p})$ then the averaged potential is
\begin{align}
\label{eq:planetpot}
 \langle \Phi_{\rm p} \rangle (r,\phi) = & -\int G \biggl\{\left[r^2 + r_{\rm p}^2 - 2rr_{\rm p}\cos(\phi-\phi_{\rm p}) \right]^{-1/2}  \nonumber \\ 
& + \frac{r\cos(\phi - \phi_{\rm p})}{r_{\rm p}^2} \biggl\} \sigma_{\rm p} \id l  ,
\end{align}
where $\sigma_{\rm p}$ is the line density for the planet and $\id l$ is the line element along the orbit. In this expression, the first term is the direct term due to the planet's gravity and the second is the indirect term due to the fact that the coordinate origin is not at the center of mass. The planet follows a Keplerian orbit with orbital elements $(\ap,\ep,\varpi_{\rm p})$. We have:
\begin{equation}
r_{p} = \frac{\ap(1-\ep^2)}{1+\ep\cos(\varpi_{\rm p}-\phi_{\rm p})} .
\end{equation}
Since equal masses are traversed in equal times, and by Kepler's second law $\id\phi_{\rm p}/\id t \propto r_{\rm p}^{-2}$, the mass element is given by
\begin{equation}
\sigma_{\rm p}\id l = \frac{\Mp r_{\rm p}^2}{\ap^2(1-\ep^2)^{1/2}}\frac{\id \phi_{\rm p}}{2\pi}.
\end{equation}
The quantity in the direct part of equation (\ref{eq:planetpot}) can be expended in terms of Laplace coefficients (see eq. [\ref{eq:lc}]) as series in the eccentricity. Evaluating the integral, we find to first order that
\begin{equation}
\begin{split}
\label{eq:avpot}
\langle \Phi_{\rm p} \rangle & = - \frac{G\Mp}{2\ap} \bigg[ \lc{1/2}{0}(\beta)  \\
& - \ep\cos(\varpi_{\rm p}-\phi)\left(1-\beta\dd{}{\beta} \right) \lc{1/2}{1}(\beta) \bigg],
\end{split}
\end{equation}
where $\beta=r/\ap$. We note that the indirect term in the potential does not contribute to this expression. Then the unperturbed (axisymmetric) potential is
\begin{equation}
\label{eq:phip}
\Phi_{\rm p} = -\frac{GM_{\rm p}}{2\ap} \lc{1/2}{0}(\beta),
\end{equation}
and the linear perturbation $\Re[\Phi_{\rm p}'(r)\me^{-\im\phi}]$ is given by
\begin{equation}
\label{eq:phipp}
\Phi_{\rm p}' = \frac{G\Mp}{2\ap}\Ep\left(1-\beta\dd{}{\beta} \right)\lc{1/2}{1}(\beta).
\end{equation}
We insert these expressions in equation (\ref{eq:dedtapp}) and simplify using identities for the Laplace coefficients and their derivatives \citep[e.g.,][]{md99} to obtain the contribution
\begin{equation}
\label{eq:appepd}
\Sigma r^2 \Omega \left(\pd{E}{t}\right)_{\rm pd} = \frac{\im G\Sigma\Mp}{4\ap}\beta \left[ \lc{3/2}{1}(\beta)E - \lc{3/2}{2}(\beta)\Ep \right],
\end{equation}
to the eccentricity equation of the disc, in agreement with equation (\ref{eq:epd1}) before softening is introduced.

In a similar way, the potential due to the continuous surface density distribution of an eccentric disc is given by
\begin{align}
\label{eq:discpot}
\langle \Phi_{\rm d} \rangle (r,\phi) = & -\int\int_{0}^{2\pi} G\Sigma(r',\phi') \\ 
&\times \left[r^2 + r'^2 - 2rr'\cos(\phi-\phi') \right]^{-1/2} r'\id r' \id \phi', \nonumber
\end{align}
where the first integral is carried over the radial extent of the disc. To evaluate the effect of self-gravity in the disc we need to insert the density distribution in the disc, taking into account the density perturbation given by equation (\ref{eq:rhop}). When this is integrated vertically the second term vanishes and we are left with an overall surface density
\begin{equation}
\Sigma(r,\phi) = \Sigma_0(r) + \Re\left[r\pd{}{r}\left(\Sigma_0 E \right)\me^{-\im\phi} \right],
\end{equation}
correct to first order. The first term of this expression contributes to the axisymmetric potential $\Phi_{\rm p,d}$, while the second contributes to the potential perturbation $\Phi_{\rm p,d}'$ in equation (\ref{eq:dedtapp}). The axisymmetric term is easy to evaluate in terms  of Laplace coefficients,
\begin{equation}
\Phi_{\rm d} = -\int\pi G\Sigma(r')\lc{1/2}{0}\left(\frac{r}{r'}\right)\id r'.
\end{equation}
The second term can be integrated by parts if the boundary conditions are such that $\rho E$ vanishes at the edges of the disc. Then we find
\begin{equation}
\Phi_{\rm d}' = \int\pi G\Sigma(r')E(r')\left(1+r'\pd{}{r'}\right)\lc{1/2}{1}\id r'.
\end{equation}
With $\beta=r/r'$ these are the continuous version of equation (\ref{eq:phip}) and (\ref{eq:phipp}), which can be recovered by setting $\Sigma(r)=(\Mp/2\pi\ap)\delta(r-\ap)$

A similar equation can be obtain for the evolution of the complex eccentricity of the planet. This can be achieved by considering once again the planet as an elliptical ring, equivalent to an annulus of the disc. In equation (\ref{eq:appepd}) the role of the annulus of disc matter at radius $r$ and the planet at distance $\ap$ can be inverted: Simply apply the transformation $(r,\Omega,\Sigma,E)\leftrightarrow(\ap,\Op,(\Mp/2\pi\ap)\delta(r-\ap),\Ep)$ to all quantities in equation (\ref{eq:appepd}), and integrate over the radial extent of the disc, to obtain equation (\ref{eq:epd2}) before smoothing is introduced.

\section{Analysis of mean-motion resonances}
\label{app:res}

A general method for deriving the rates of change of the complex eccentricities of the disc and planet due to a mean-motion resonance was derived in \citet{ogilvie07}. From equations (99) and (101) of that paper we find that the local growth rates due to an interior $j:j-2$ eccentric Lindblad resonance (ELR) with $j\geq 3$, correct to first order in eccentricity, are
\begin{equation}
\begin{split}
\left(\pd{E}{t} \right)_{\rm ELR} = & \frac{1}{r^2\Omega} \frac{G\Mp^2}{M_*} \frac{2\pi}{3(j-2)} \frac{r^2}{\ap^2} \\
& \times 2 f_{45} (2f_{45}E+f_{49}\Ep)\delta(r-\rres),
\end{split}
\end{equation}
\begin{equation}
\begin{split}
\left( \dd{\Ep}{t} \right)_{\rm ELR} = & \frac{1}{\ap^2\Op} \frac{G\Mp}{M_*} 2\pi r\Sigma \frac{2\pi}{3(j-2)} \frac{r^2}{\ap^2} \\
& \times f_{49} (2f_{45}E+f_{49}\Ep)\bigg|_{r=\rres},\end{split}
\end{equation}
where
\begin{equation}
f_{45} = \frac{1}{8}\left[(-5j+4j^2) + (-2+4j)\beta\dd{}{\beta} + \beta^2\dd{^2}{\beta^2} \right]\lc{1/2}{j}(\beta),
\end{equation}
\begin{equation}
f_{49} = \frac{1}{4}\left[(-2+6j-4j^2) + (2-4j)\beta\dd{}{\beta} - \beta^2\dd{^2}{\beta^2} \right]\lc{1/2}{j-1}(\beta),
\end{equation}
evaluated at $\beta=\rres/\ap = [(j-2)/j]^{2/3}$, are dimensionless coefficients from the expansion of the disturbing function given by \citet{md99}. These expressions are of the form of equations (\ref{eq:elrd}) and (\ref{eq:elrp}) with
\begin{equation}
\begin{split}
\mathscr{A} & = \left[\frac{2\pi}{3(j-2)}\right]^{1/2}  2 \beta f_{45},  \\
\mathscr{B} & = -\left[\frac{2\pi}{3(j-2)} \right]^{1/2}\beta f_{49},
\end{split}
\end{equation}
In Section \ref{sec:res} we have also assumed that the resonances have a non-zero width, which we modelled by replacing the $\delta$ function by an off-centred Gaussian profile.

The equivalent result for an exterior $j:j-2$ ELR is
\begin{equation}
\begin{split}
\label{eq:appab}
\mathscr{A} & = \left(\frac{2\pi}{3j}\right)^{1/2}  2 \left(f_{53} - \frac{3}{8\beta^2} \delta_{j,3} \right), \\
\mathscr{B} & = -\left(\frac{2\pi}{3j} \right)^{1/2}f_{49},
\end{split}
\end{equation}
with
\begin{equation}
f = \frac{1}{8}\left[ (2-7j+4j^2) + (-2+4j)\beta\dd{}{\beta} + \beta^2\dd{^2}{\beta^2}  \right] \lc{1/2}{j-2}(\beta),
\end{equation}
evaluated at $\beta=\ap/\rres=[(j-2)/j]^{2/3}$. 

For the eccentric corotation resonances (ECRs) we use equations (112) and (113) of \citet{ogilvie07}. The effect is proportional to the local vortencity gradient of the disc, and we obtain by a similar method equations (\ref{eq:ecrd}) and (\ref{eq:ecrp}) without the Gaussian spreading. For an interior $j:j-1$ ECR with $j\geq 2$,
\begin{equation}
\begin{split}
\mathscr{C} & = -\left(\frac{2\pi}{3}\right)^{1/2} \beta f_{27}, \\
\mathscr{D} & = \left(\frac{2\pi}{3}\right)^{1/2} \beta (f_{31}-2\beta\delta_{j,2}),
\end{split}
\end{equation}
where 
\begin{equation}
f_{27} = \frac{1}{2}\left(-2j - \beta\dd{}{\beta}\right) \lc{1/2}{j}(\beta),
\end{equation}
\begin{equation}
f_{31} = \frac{1}{2}\left[ (-1+2j) + \beta\dd{}{\beta} \right] \lc{1/2}{j-1}(\beta),
\end{equation}
evaluated at $\beta=\rres/\ap = [(j-1)/j]^{2/3}$. For an exterior $j-1:j$ ECR,
\begin{equation}
\begin{split}
\mathscr{C} & = -\left(\frac{2\pi}{3}\right)^{1/2} f_{27}, \\
\mathscr{D} & = \left(\frac{2\pi}{3}\right)^{1/2} (f_{31}-\frac{1}{2\beta^2}\delta_{j,2}),
\end{split}
\end{equation}
with $\beta=\ap/\rres = [(j-1)/j]^{2/3}$.

We obtain a correspondence with \citet{ward88} by noting that in the limit of large $j$,
\begin{equation}
\mathscr{A}\sim\mathscr{B}\sim \left[5K_{0}\left(\frac{4}{3}\right) + \frac{19}{4} K_1 \left(\frac{4}{3}\right) \right] \left( \frac{j^3}{6\pi}\right)^{1/2} = 0.6941j^{2/3},
\end{equation}
\begin{equation}
\mathscr{C}\sim\mathscr{D}\sim \left[2K_{0}\left(\frac{2}{3}\right) + K_1 \left(\frac{2}{3}\right) \right] \left( \frac{2 j^2}{3\pi}\right)^{1/2} = 1.1606j,
\end{equation}
where $K$ is the modified Bessel function. The competition between ELRs and ECRs described by \citet{gt80} is such that $(3/8)\mathscr{D}^2/j^2$ exceeds $\mathscr{B}^2/j^3$ by approximatively five per cent in the limit of large $j$.

\section{Discretization of linearized equations}
\label{app:discret}

\subsection{Adiabatic model}

\subsubsection{General equation}

The general form of the eccentricity evolution equation for an inviscid, non-self-gravitating disc can be written as, in the adiabatic model:
\begin{equation}
-\im\Sigma r^2 \Omega \pd{E}{t} = \frac{1}{r}\pd{}{r}\left(\ff\pd{E}{r}\right) + \gf E,
\label{eq:sladia}
\end{equation}
where $\Sigma(r)$ is the surface density, $\Omega(r)$ is the Keplerian angular velocity. The functions $\ff$ and $\gf$ are real functions of $r$, which read for each model:
\begin{itemize}
\item 2D adiabatic model:
\begin{equation}
\ff = \frac{1}{2}\gamma Pr^3.
\end{equation}
\begin{equation}
\gf = \frac{r}{2}\dd{P}{r}.
\end{equation}
\item 3D adiabatic model:
\begin{equation}
\ff = \frac{1}{2}\left(2-\frac{1}{\gamma}\right)\gamma Pr^3.
\end{equation}
\begin{equation}
\gf = \frac{1}{2}\left(4-\frac{3}{\gamma}\right)r\dd{P}{r} + \frac{3}{2}\left(1+\frac{1}{\gamma}\right)P.
\end{equation}

When viscosity is included, $\ff$ becomes a complex function. Resonances, self-gravity and gravitational interactions with a planet simply add real or complex terms to $\gf$. Overall, the problem takes the form of a Sturm-Liouville equation, with possibly complex functions of radius.

\subsubsection{Continuous model}

For a disc with free boundaries at $r=\rin$ and $r=\rout$, the boundary conditions are 
\begin{equation}
\ff\pd{E}{r} = 0 \quad \text{at} \qquad r=\rin \quad \text{and} \quad r=\rout.
\end{equation}
These conditions are automatically satisfied if the model has $\ff(\rin)=\ff(\rout)=0$ and if the solution is regular at the boundaries. Otherwise they correspond to $\partial{E}/\partial r =0$.

The angular momentum deficit
\begin{equation}
A = \int_{\rin}^{\rout} \frac{1}{2}\vert E \vert^2 \Sigma r^2\Omega2\pi r \id r
\end{equation}
is then conserved. Indeed, if we denote by $c.c.$ the complex conjugate: 
\begin{equation}
\begin{split}
\dd{A}{t} &= \int_{\rin}^{\rout} \frac{1}{2}\left(E^* \pd{E}{t} + \text{c.c.} \right) \Sigma r^2 \Omega 2\pi r \id r\\
&= \int_{\rin}^{\rout} \im \pi E^* \left[ \pd{}{r}\left( \ff \pd{E}{r} \right) + GrE \right] \id r + \text{c.c.}\\
&= -\int_{\rin}^{\rout} \im \pi \ff \left|\pd{E}{r}\right|^2 \id r + \text{c.c.} \\
&= 0. 
\end{split}
\end{equation}

For normal modes $\propto \me^{\im\omega t}$ we have the eigenvalue problem
\begin{equation}
\omega\Sigma r^2\Omega E = \frac{1}{r}\dd{}{r}\left(\ff\dd{E}{r}\right) + \gf E,
\end{equation}
with the associated integral relation
\begin{equation}
\label{eq:intadia}
\omega\int_{\rin}^{\rout} \Sigma r^2\Omega \left|E\right|^2 r \id r = \int_{\rin}^{\rout} \left(- \ff \left|\pd{E}{r}\right|^2  + r\gf \left|E\right|^2  \right)\id r.
\end{equation}

\subsubsection{Discrete model}
We split the disc into $n$ annuli with $i=1,2,\ldots , n$, such that annulus $i$ occupies the interval $r_{i-1}<r<r_i$, with $r_0=\rin$ and $r_n=\rout$. Multiplying eq. (\ref{eq:sladia}) by $2\pi r$ and integrate from $r_{i-1}$ to $r_i$, we have:
\begin{equation}
\begin{split}
-\im\dd{}{t}\int_{r_{i-1}}^{r_i}\Sigma r^2\Omega E 2\pi r\id r & = 2\pi \left[\ff \pd{E}{r}\right]_{r_{i-1}}^{r_i} + \int_{r_{i-1}}^{r_i} \gf E 2\pi r\id r.
\end{split}
\end{equation}
We associate a single eccentricity $E_i$ with annulus $i$ and rewrite this equation as:
\begin{equation}
-\im J_i E_i = 2\pi\left[\ff \pd{E}{r} \right]_{r_{i-1}}^{r_i} + g_i J_i E_i
\end{equation}
where
\begin{equation}
J_i = \int_{r_{i-1}}^{r_i} \Sigma r^2\Omega 2\pi r \id r,
\end{equation}
\begin{equation}
g_i J_i = \int_{r_{i-1}}^{r_i} \gf 2\pi r \id r,
\end{equation}

The discrete eccentricity equation is then:
\begin{equation}
\begin{split}
\label{eq:discretizeadia}
-\im J_i \dd{E_i}{t} & = g_i J_i E_i \\ 
& + 2\pi\left[ \frac{\ff_i E_{i+1}}{\delta r_i} - \frac{\ff_i E_{i}}{\delta r_i} - \frac{\ff_{i-1} E_{i}}{\delta r_{i-1}} + \frac{\ff_{i-1} E_{i-1}}{\delta r_{i-1}}
 \right].
\end{split}
\end{equation}
Note that $\ff_0 = \ff_n =0$ because of the boundary conditions. 

For normal modes $\propto \me^{\im\omega t}$, we multiply eq. (\ref{eq:discretizeadia}) by the complex conjugate of the eccentricity, and sum over $i$ to get
\begin{equation}
\omega \sum_{i=1}^n J_i \left|E_i\right|^2 = \sum_{i=1}^n g_i J_i \left|E_i\right|^2
 - 2\pi \sum_{i=1}^n \frac{F_i}{\delta r_i} \left| E_{i+1} - E_i \right|^2,
\end{equation}
where we have rearranged the summation indices in the last term on the right-hand side. This is the discrete equivalent of equation (\ref{eq:intadia}), and can be used to computed the integral quantities of Section \ref{sec:int}.

For inviscid discs without planets or self-gravity, the AMD is then conserved in the following form:
\begin{equation}
\dd{}{t}\left(\sum_{i=1}^n\frac{1}{2}|E_i|^2J_i \right) = 0. 
\end{equation}

\subsection{Locally isothermal model}

\subsubsection{General equation}

For an inviscid disc, without self-gravity and interactions with a planet, the isothermal equation described in Section \ref{sec:pres} can be written in the general form:
\begin{equation}
\label{eq:isoSL}
-\im\Sigma r^2 \Omega \cs \pd{}{t}\left(\frac{E}{\cs}\right) = \frac{1}{r}\pd{}{r}\left(\ff\pd{}{r}\left(\frac{E}{\cs}\right)\right) + \gf\cs\frac{E}{\cs},
\end{equation}
where we have assumed that $\cs$ does not depend on time. The isothermal model has the form of a Sturm-Liouville problem for the function $E/\cs$. Again, $\ff$ and $\gf$ are real functions of $r$, which read for each model:
\item 2D locally isothermal model:
\begin{equation}
\ff = \frac{1}{2}\Sigma c_{\rm s}^4 r^3.
\end{equation}
\begin{equation}
\gf = \frac{r}{2}\dd{}{r}\left(\Sigma\cs \right)
\end{equation}
\item 3D locally isothermal model:
\begin{equation}
\ff = \frac{1}{2}\Sigma c_{\rm s}^4 r^3.
\end{equation}
\begin{equation}
\gf = \frac{r}{2}\dd{}{r}\left(\Sigma\cs \right) + \frac{3}{2r}\Sigma\dd{}{r}\left(\cs r^2\right).
\end{equation}
\end{itemize}
Note that in the isothermal case, the definition of the function $\ff$ differs from that of the adiabatic case by a factor of $\cs$. Similarly to the adiabatic case, resonances, interactions with the planet and self-gravity can be included in the $\gf$ term. However, viscosity requires a special treatment, as we highlight below. 

\subsubsection{Continuous model}

In the locally isothermal model, the AMD is not conserved. However, following the same steps that in the adiabatic case, it is straightforward to show that the quantity
\begin{equation}
\int_{\rin}^{\rout} \frac{1}{2}\vert E \vert^2 \frac{\Sigma r^2\Omega}{\cs}2\pi r \id r
\end{equation}
is conserved.

For normal modes $\propto \me^{\im\omega t}$ we have the eigenvalue problem with the associated integral relation:
\begin{equation}
\begin{split}
\label{eq:intiso}
\omega\int_{\rin}^{\rout} \frac{\Sigma r^2\Omega}{\cs} \left|E\right|^2 r \id r & = \int_{\rin}^{\rout} \left(- \ff \left|\pd{}{r}\left(\frac{E}{\cs}\right)\right|^2  \right. \\
& \left. + \frac{r\gf}{\cs} \left|E\right|^2  \right)\id r.
\end{split}
\end{equation}
In order to derive the integral relations of Section \ref{sec:int}, it is more convenient to consider the following integral relation, in which the AMD appears explicitly:
\begin{equation}
\begin{split}
\omega\int_{\rin}^{\rout} \Sigma r^2\Omega \left|E\right|^2 r \id r & = \int_{\rin}^{\rout}\Bigg( \frac{\ff}{c_{\rm s}^4}\dd{\cs}{r} E\pd{E^*}{r} \\
&  - \frac{\ff}{\cs}\left|\pd{E}{r}\right|^2  + r\gf \left|E\right|^2  \Bigg) \id r.
\end{split}
\end{equation}

\subsubsection{Discrete model}

Following the same steps as in the adiabatic case, we find
\begin{equation}
\begin{split}
-\im J_i \dd{E_i}{t} & = g_i J_i E_i + 2\pi\left[ \frac{\ff_i}{\delta r_i}\frac{ E_{i+1}}{c_{{\rm s}, i+1}^2} - \frac{\ff_i}{\delta r_i}\frac{ E_i}{c_{{\rm s}, i}^2} \right. \\
& \left. - \frac{\ff_{i-1}}{\delta r_{i-1}} \frac{E_i} {c_{{\rm s}, i}^2} + \frac{\ff_{i-1}}{\delta r_{i-1}}\frac{ E_{i-1}}{c_{{\rm s}, i-1}^2}
 \right].
\end{split}
\end{equation}

As in the adiabatic case, resonances, self-gravity and gravitational interactions with a planet simply add real or complex terms to $\gf$, and the problem can still be formulated as a Sturm-Liouville equation. It is worth noting that when viscosity is included, it is not possible to formulate the problem in the form of a Sturm-Liouville equation. The locally isothermal discretized model with viscosity reads
\begin{equation}
\label{eq:discretizeiso}
\begin{split}
& -\im J_i \dd{E_i}{t} = g_i J_i E_i \\
 & + 2\pi\left[ \frac{\ff_i}{\delta r_i}\frac{ E_{i+1}}{c_{{\rm s}, i+1}^2} - \frac{\ff_i}{\delta r_i}\frac{ E_i}{c_{{\rm s}, i}^2} - \frac{\ff_{i-1}}{\delta r_{i-1}} \frac{E_i} {c_{{\rm s}, i}^2} + \frac{\ff_{i-1}}{\delta r_{i-1}}\frac{ E_{i-1}}{c_{{\rm s}, i-1}^2}
 \right]. \\
 & -2\pi\im\ab \left[ \frac{\ff_i}{c_{{\rm s}, i}^2}\frac{ E_{i+1}}{\delta r_i} - \frac{\ff_i}{c_{{\rm s}, i}^2}\frac{ E_i}{\delta r_i} - \frac{\ff_{i-1}}{c_{{\rm s}, i-1}^2} \frac{E_i} {\delta r_{i-1}} + \frac{\ff_{i-1}}{c_{{\rm s}, i-1}^2}\frac{ E_{i-1}} {\delta r_{i-1}}
 \right].
 \end{split}
\end{equation}

For normal modes $\propto \me^{\im\omega t}$, we multiply eq. (\ref{eq:discretizeiso}) by the complex conjugate of the eccentricity divided by the sound speed square, and sum over $i$ to get
\begin{equation}
\begin{split}
& \omega \sum_{i=1}^n J_i \frac{\left|E_i\right|^2}{c_{{\rm s}, i}^2}  = \sum_{i=1}^n g_i J_i\frac{\left|E_i\right|^2}{c_{{\rm s}, i}^2} 
  - 2\pi\sum_{i=1}^n \frac{F_i}{\delta r_i} \left| \frac{E_{i+1}}{c_{{\rm s}, i+1}^2} - \frac{E_i}{c_{{\rm s}, i+1}^2} \right|^2  \\
& + 2\pi \im\ab  \sum_{i=1}^n \frac{F_i}{c_{{\rm s},i}^2\delta r_i} \left( \frac{|E_{i+1}|^2}{c_{{\rm s},i+1}^2} - \frac{E_{i+1}E_i^*}{c_{{\rm s},i}^2} - \frac{E_iE_{i+1}^*}{c_{{\rm s},i+1}^2} + \frac{|E_i|^2}{c_{{\rm s},i}^2}\right).
 \end{split}
\end{equation}

Note that the last term has both a real and imaginary part. This equation is the discrete version of equation (\ref{eq:intadia}). It provides an alternative way of computing the precession rate and growth rate compared with the integral relations we derived in Section \ref{sec:int}, but both methods give equivalent results. 

\section{Schr\"odinger equation for eccentric discs}
\label{app:schrod}

In this appendix we show how the results of Section \ref{sec:schrod} can be generalized, in particular to the case where $H/r$ is no longer a constant.

The equations of the 2D and 3D adiabatic and locally isothermal models
for an eccentric normal mode $\propto\me^{\im\omega t}$ can be
written in the form
\begin{equation}
\label{eq:generalSL}
  \omega AE=\dd{}{r}\left(B\dd{E}{r}\right) + CE.
\end{equation}
where $A(r)$, $B(r)$ and $C(r)$ are real functions. In the adiabatic case, the transformation from Equations (\ref{eq:2dadia}) and (\ref{eq:3dadia}) to Equation (\ref{eq:generalSL}) is straightforward and we have:
\begin{equation}
  A=2\Sigma r^3\Omega
\end{equation}
\begin{equation}
  B=aPr^3,
\end{equation}
\begin{equation}
  C=br^2\dd{P}{r} + crP
\end{equation}
In 2D we have $a=\gamma$, $b=1$ and $c=0$. In 3D we have $a=2-\gamma^{-1}$, $b=4-3\gamma^{-1}$ and $c=3(1+\gamma^{-1})$.

In the isothermal case the transformation is achieved by multiplying Equation (\ref{eq:2diso}) or (\ref{eq:3diso}) by $2r/\cs$ to get Equation (\ref{eq:generalSL}) with
\begin{equation}
  A=\frac{2\Sigma r^3\Omega}{\cs}
\end{equation}
\begin{equation}
  B=\Sigma r^3,
\end{equation}
\begin{equation}
  C=\frac{r^2}{\cs}\dd{}{r}\left(\Sigma\cs\right) - \frac{1}{\cs}\dd{}{r}\left(\Sigma r^3 \dd{\cs}{r}  \right) + \frac{3\delta_{\rm 3D}}{\cs}\Sigma\dd{}{r}\left(\cs r^2\right),
\end{equation}
where $\delta_{\rm 3D}=0$ in 2D and $\delta_{\rm 3D}=1$ in 3D. We note that the way we have written the isothermal evolution equation here is different from the one we used for the discretization method (Eq. \ref{eq:isoSL}), but both formulations are equivalent.

In order to rewrite Equation (\ref{eq:generalSL}) into a Schr\"odinger equation, we transform $r$ and $E$ to a new independent variable $x(r)$ and a new dependent
variable $\Psi(x)$ such that
\begin{equation}
 \dd{x}{r} = \left( \frac{\omega_0 A}{B}\right)^{1/2}
\end{equation}
\begin{equation}
  E=\frac{\Psi}{f},
\end{equation}
with
\begin{equation}
  f=(AB)^{1/4},
\end{equation}
where $\omega_0$ is an arbitrary positive constant with the dimensions
of frequency. Then we obtain the dimensionless time-independent
Schr\"odinger equation
\begin{equation}
  -\dd{^2\Psi}{x^2}+V\Psi=\mathcal{E}\Psi
\end{equation}
with an effective energy eigenvalue
\begin{equation}
  \mathcal{E}=-\frac{\omega}{\omega_0}
\end{equation}
and an effective potential $V(x)$ given by
\begin{equation}
\label{eq:vschrod}
  V=-\frac{C}{\omega_0A}+\frac{1}{f}\dd{^2f}{x^2}.
\end{equation}

We then note that, both in the adiabatic and isothermal cases, we have
\begin{equation}
  \frac{A}{B}=\frac{2\Sigma\Omega}{\gamma_1P}=\frac{2}{\gamma_1H^2\Omega},
\end{equation}
where $\gamma_1=a$ for the adiabatic models and $\gamma_1=1$ for the locally isothermal models. Given that
\begin{equation}
  \Omega=\Omega_\mathrm{in}\left(\frac{r}{r_\mathrm{in}}\right)^{-3/2}
\end{equation}
and
\begin{equation}
  \frac{H}{r}=h_0\left(1-\sqrt{\frac{r_\mathrm{in}}{r}}\right)^{2/9},
\end{equation}
where $h_0=\mathrm{constant}$, we choose
\begin{equation}
  \omega_0=\frac{\gamma_1}{2}h_0^2\Omega_\mathrm{in},
\end{equation}
so that
\begin{equation}
  \dd{x}{r}=\frac{1}{r}\left(\frac{r}{r_\mathrm{in}}\right)^{3/4}\left(1-\sqrt{\frac{r_\mathrm{in}}{r}}\right)^{-2/9}.
\end{equation}
This equation can be integrated to give:
\begin{equation}
  x=\frac{4}{3}\left(\frac{r}{r_\mathrm{in}}\right)^{3/4}\,_2F_1\left(-\frac{3}{2},\frac{2}{9};-\frac{1}{2};\sqrt{\frac{r_\mathrm{in}}{r}}\right),
\end{equation}
where $_2F_1$ is a hypergeometric function. More generally, in a model in which
\begin{equation}
  \frac{H}{r}=h_0\left(\frac{r}{r_\mathrm{in}}\right)^p\left(1-\sqrt{\frac{r_\mathrm{in}}{r}}\right)^{q},
\end{equation}
we have (with a suitable choice of $\omega_0$)
\begin{equation}
  x=\frac{4}{3-4p}\left(\frac{r}{r_\mathrm{in}}\right)^{(3/4)-p}\,_2F_1\left(-\frac{3}{2}+2p,q;-\frac{1}{2}+2p;\sqrt{\frac{r_\mathrm{in}}{r}}\right).
\end{equation}

The solution presented in this appendix is a generalisation of that of Section \ref{sec:schrod}. The effect of a planet or short-range forces can easily be included in Equation (\ref{eq:vschrod}). However we find that the solution we present here does not significantly affect the shape of the effective potential well, and hence the physical interpretation remains the same.

\end{document}